\documentclass[preprint1]{aastex}

\slugcomment{To appear in the Astronomical Journal in 2008}

\usepackage{palatcm, color}
\usepackage{graphicx}

\begin{document}

\shorttitle{Atlas of Orion Circumstellar Disks}
\shortauthors{Ricci et al.}

\title{The HST/ACS Atlas of Protoplanetary Disks  in the Great Orion Nebula\footnote{Based on observations made with the NASA/ESA Hubble Space Telescope, obtained at the Space Telescope Science Institute, which is operated by the Association of Universities for Research in Astronomy, Inc., under NASA contract NAS 5-26555. These observations are associated with program 10246.}}



\author{L. Ricci\altaffilmark{2}, M. Robberto and D. R. Soderblom}
\affil{Space Telescope Science Institute, 3700 San Martin Dr., Baltimore MD 21218}


\altaffiltext{2}{present address: European Southern Observatory,
Garching bei M\"unchen, Germany}

\email{robberto@stsci.edu}




\begin{abstract}

We present the atlas of protoplanetary disks in the Orion Nebula
based on the ACS/WFC images obtained for the \emph{HST Treasury
Program on the Orion Nebula Cluster}. The observations have been
carried out in 5 photometric filters nearly equivalent to the
standard $B, V, H\alpha, I,$ and $ z$ passbands. Our master
catalog lists 178 externally ionized proto-planetary disks
(\emph{proplyds}), 28 disks seen only in absorption against the
bright nebular background (\emph{silhouette disks}, 8 disks seen
only as dark lanes at the midplane of extended polar emission
(\emph{bipolar nebulae} or \emph{reflection nebulae}) and 5
sources showing jet emission with no evidence of neither external
ionized gas emission nor dark silhouette disks. Many of these
disks are associated with jets seen in $H\alpha$ and circumstellar
material detected through reflection emission in our broad-band
filters; approximately 2/3 have identified counterparts in x-rays.
A total of 47 objects (29 proplyds, 7 silhouette disks, 6 bipolar
nebulae, 5 jets with no evidence of proplyd emission or silhouette
disk) are new detections with HST. We include in our list 4
objects previously reported as circumstellar disks which have not
been detected in our HST/ACS images either because they are hidden
by the bleeding trails of a nearby saturated bright star or
because of their location  out of the HST/ACS Treasury Program
field. Other 31 sources previously reported as extended objects do
not harbor a stellar source in our HST/ACS images. We also report
on the detection  of 16 red, elongated sources. Their location at
the edges of the field, far from the Trapezium Cluster core
($\gtrsim 10'$), suggests that these are probably background
galaxies observed through low extinction regions of the Orion
Molecular Cloud OMC-1.

\end{abstract}





\section{Introduction}

The Orion Nebula (M42, NGC 1976) is a unique laboratory for
studying the physical processes related to star and planet
formation. It harbors one of the richest and youngest clusters
(Orion Nebula Cluster, ONC)\ in the solar neighborhood, spanning
the full spectrum of stellar and sub-stellar masses down to a few
Jupiter masses \citep{Luca00}. In 1979 several compact
photoionized knots were firstly detected in the central region of
the Orion Nebula as emission-line sources \citep{Laqu79}, and then
important follow-up studies were made in radio \citep{Gara87,
Chur87} and via emission-line spectroscopy \citep{Meab88, Meab93,
Mass93}. Since the early 1990's, \emph{Hubble Space Telescope}
(HST) observations of the ONC have been fundamental for clarifying
the main characteristics of these young stellar objects (YSO) and
their accretion disks. After the pioneering surveys of
\citet{Odel93} and \citet{Pros94}, performed with the
spherically-aberrated WF/PC, \citet{Odel94} used WFPC2 to discover
several externally ionized proto-planetary disks
(\emph{proplyds}), as well as a number of disks seen only in
absorption against the bright nebular background (\emph{silhouette
disks}), both rendered visible by their location in or near the
core of the \ion{H}{2} region. Following this discovery, other HST
programs have increased the number of known objects \citep{Odel96,
McCa96, Ball98, Ball00}. \citet{Odel01} and \citet{Smit05},
targeting areas out of the core, showed that these systems are
ubiquitous across the Great Orion Nebula.

So far a total of $\sim$ 200 silhouette disks and bright proplyds
has been revealed by the HST observations of the Orion Nebula, the
large majority through narrow-band filters centered on the
H$\alpha$ $\lambda 6563$ emission lines, and occasionally through
filters centered on the [N II] $\lambda 6583$, [O I] $\lambda
6300$, [O III] $\lambda 5007$ and [S II] $\lambda 6717+6731$
lines.

In this paper we present an atlas of multi-color observations of
circumstellar disks and resolved circumstellar emission obtained
with the Wide Field Channel of the Advanced Camera for Surveys
(ACS/WFC). These images are part of the \emph{HST Treasury Program
on the Orion Nebula Cluster} (Cycle 13, GO Program 10246, P.I. M.
Robberto), aimed at measuring with high precision the main stellar
parameters of the cluster members. For this reason, the Treasury
Program used broad-band filters to obtain the most accurate
photometry of each source, together with H$\alpha$ narrow-band
images to address the presence of circumstellar emission that may
contaminate the photometry and the point spread function of the
broad-band data. The combination of broad-band and narrow-band
images opens a new window on the study of disks in the OMC. It
also makes it possible to detect disks where the nebular background is too
faint, thanks to the light of the central stars reflected by the
circumstellar material at the disk's polar regions (reflection
nebulae).

After a brief description of the observations (\S 2), we present
the new ACS/WFC images of all circumstellar disks  (\S 3). We then
provide a complete catalogue of circumstellar disks in the Orion
Nebula, including also the few disks that were not detected in our
programs for a variety of reasons (\S 4). Finally, after a brief
description of the new proplyds, silhouette disks, and bipolar
nebulae (\S 5, \S 6) we present the images of 16 red, elongated,
and diffuse objects which most probably represent galaxies seen
through the background curtain provided by the Orion Molecular
Cloud (OMC-1, \S 7). A few remarkable objects are being
investigated and will be discussed in separate papers (see e.g.
\citet{Robb08b} on 124-132).

\section{Observations}

The images  have been extracted from the large dataset (520
images)\ of ACS/WFC observations  executed between November 2004
and April 2005. The ACS/WFC survey has covered an area of about
450 square arcmin, centered about 4 arcmin southwest
of the Trapezium Cluster. The filters and exposure times are
listed in Table \ref{Tab:Filters}. The narrow-band F658N filter
transmits both H$\alpha ~ \lambda$6583 and [N II]
$\lambda$6583, but it is conventionally referred to as the
ACS H$\alpha$ filter. Due to the
 dithering strategy adopted for the survey, most of the field has been
exposed two times (or more, occasionally)\ so the total
integration time is typically twice that reported in
Table~\ref{Tab:Filters}. Only at the edges of the ACS survey field
was a single image obtained.  Images have been combined using the
Py-Drizzle algorithm, allowing for removal of cosmic rays when two
or more exposures were available. Sources lying at the outer edges
of the nebula are therefore clearly recognizable by the presence
of uncorrected cosmic rays in our images. The mosaics of ACS\
images have been registered against the  2MASS catalog
\citep{Cutr03} to derive absolute astrometry of the sources
accurate to approximately 1/2 pixel (25~milliarcsec).

Each individual ACS image, both raw and drizzled, has been
visually inspected for source identification, resulting in a
master catalogue of $\sim$3,200 stellar or compact sources. Each
source has been classified as either a single star, binary,
photoevaporated disk, dark silhouette, or candidate galaxy. The
last three classes constitute the sample presented in this paper.
Note that since the ACS\ catalog targets stellar sources, it does
not include Herbig-Haro objects, bow-shocks and jets unless they
are closely associated (within $\approx 1''$) with a point source.
Further details on the \emph{HST Treasury Program}  observing
strategy and on data-reduction procedures are given in
\citet{Robb08}, whereas the complete photometric catalog is given
in \citet{Sode08}.

\section{New HST/ACS Images of Circumstellar Disks}

We found 219\ sources that show distinct evidence of circumstellar
matter. Of them, 178\ are externally ionized protoplanetary disks
seen in emission, 5 show jet emission in $H\alpha$ with no
evidence of neither external ionized gas emission nor dark
silhouette disk, 36\ can be classified as dark silhouette disks.
They are directly visible either in absorption against the nebular
background or revealed through the blocking of light coming from
their central star or by the presence of detached bipolar lobes.

Figures \ref{fig_first_ppd}$-$\ref{fig_jet} show the ACS/WFC
images of our sources in the 5 photometric bands used for our ACS
observations. In particular, the ionized protoplanetary disks seen
in emission are reported in Figures
\ref{fig_first_ppd}$-$\ref{fig_last_ppd}, while Figures
\ref{fig_first_sd}$-$\ref{fig_last_sd} show the dark silhouette
disks, Figure \ref{fig_rn} the reflection nebulae and Figure
\ref{fig_jet} the jets with no external ionized gas emission or
silhouette disk. Each frame is $100 \times 100$ ACS/WFC pixels,
corresponding to $\sim 5'' \times 5''$, or $\sim 2000 \times 2000$
AU at the distance of the Orion Nebula, here assumed to be $\sim$
420 pc (Menten\ et al. 2007).

In each row, we report the images in the five photometric bands,
in order of increasing wavelength: F435W, F555W, F658N,
F775W, F850LP. The grey scale goes from 2$\sigma$ below the
average sky level through 3$\sigma$ above it, where both the
average and $\sigma$ have been estimated using an iterative algorithm
to reject outliers. The color images at the end of each row were
created in this way: the intensity of blue is the
average of the fluxes measured in the F435W and F555W bands,
the intensity of red is the average of F775W and F850LP,
and the intensity of green is the flux measured in the F658N filter only.

All the FITS files from which these images have been taken are in
the electronic version of the \emph{Astronomical Journal}. In
these drizzled images the pixel values are in counts per seconds,
and an estimate in magnitudes of the photometry of a source can be
directly extracted by

\begin{equation}
  m_X=-2.5\cdot Log \ F_X + ZP_X,
\end{equation}
where $X$ is the passband of interest, $F_X$ is the observed flux
of the source in counts per seconds in the passband $X$, and
$ZP_X$ is the zero-point magnitude in the passband $X$ for a
certain photometric system. In Table \ref{Tab:Zero_Points} we list
the zero-point magnitudes derived by the Photometric Calibration
of the HST/ACS camera \citep{Siri05} in the VEGAMAG, ABMAG and
STMAG standard photometric systems for the filters used by our
survey.

\section{The Catalog of Circumstellar Disks in Orion}

We have searched the original literature and the
Simbad\footnote{http://simbad.u-strasbg.fr/simbad} database to
cross-identify each object that we found. For 170 of them there is
a previous HST\ classification as proplyd, silhouette disk or
compact non-stellar object. In particular, our catalogue includes
all the sources identified by \citet{Odel94, Odel96, Ball00,
Odel01, Smit05}. For 34 of these objects we could not confirm
their nature as proplyds or silhouette disks. Our observations
missed 3 sources (158-314, 163-322, 163-323), hidden by the
bleeding trail of saturated bright stars, and the silhouette disk
216-0939 \citep{Smit05} which is located outside the field covered
by the HST Treasury Program. We have excluded these sources from
our main catalogue of disks, listing them separately in Table
\ref{other_disks_tab}. Regarding the other 30 sources the HST/ACS
images (Figures \ref{fig_first_other}$-$\ref{fig_last_other})\
show that some are close binary systems with no visible
circumstellar emission, some are Herbig-Haro objects. We have
listed these objects in Table \ref{OW_BOM_objects_tab}, in which
column (11) points out the objects type as it appears from our
images. However it is important to note that some of these objects
may still have low ionization circumstellar emission (e.g. from
[OIII] emission line), since the ACS filters would not pick this
up. On the other hand, our images provide the first identification
for 63 objects, of which 29 are proplyds, 7 are silhouette disks,
6 are bipolar nebulae, 5 are jets with no external ionized gas
emission or silhouette disk, and other 16 are probably galaxies.

\begin{figure*}
\plottwo{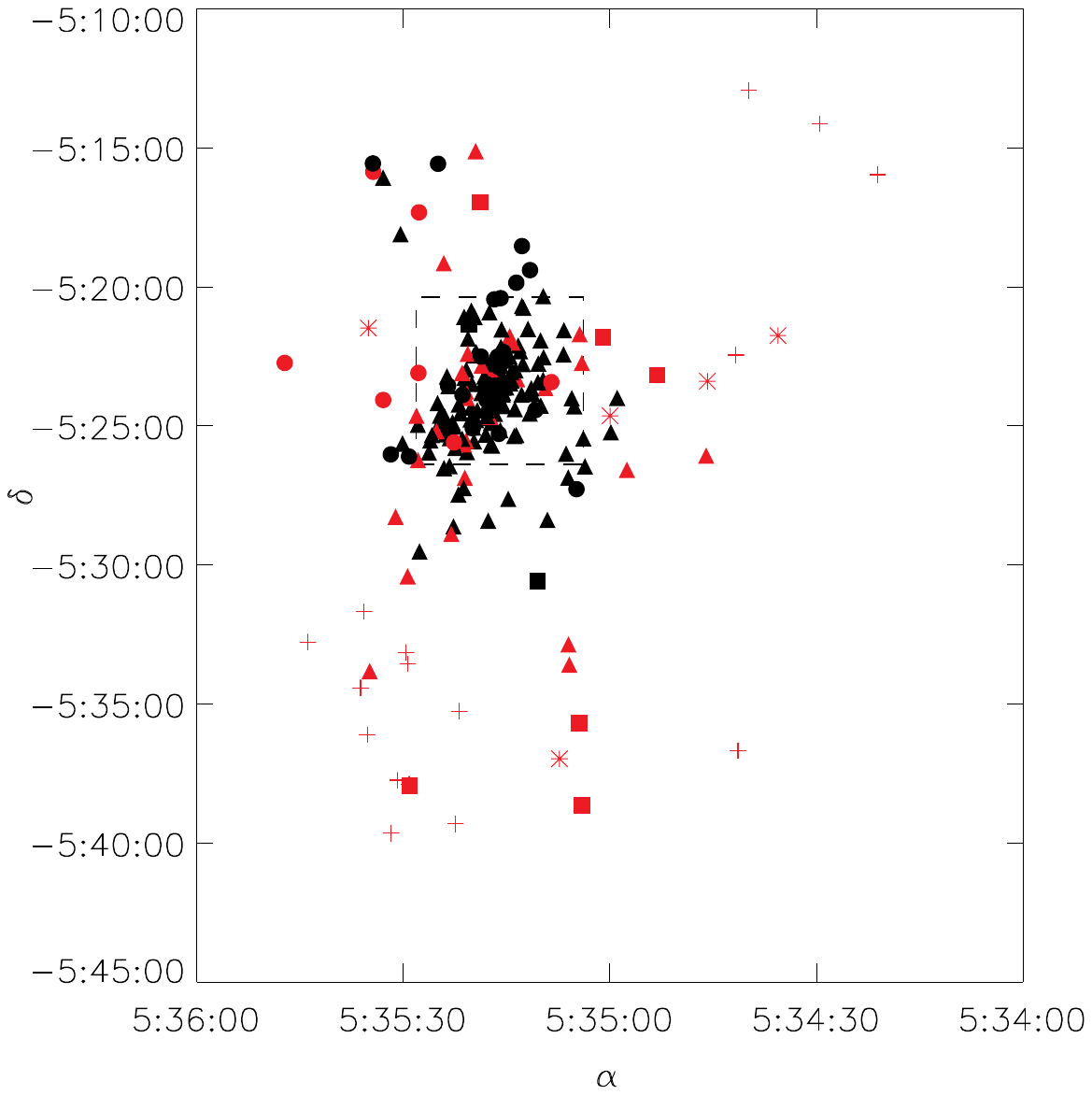}{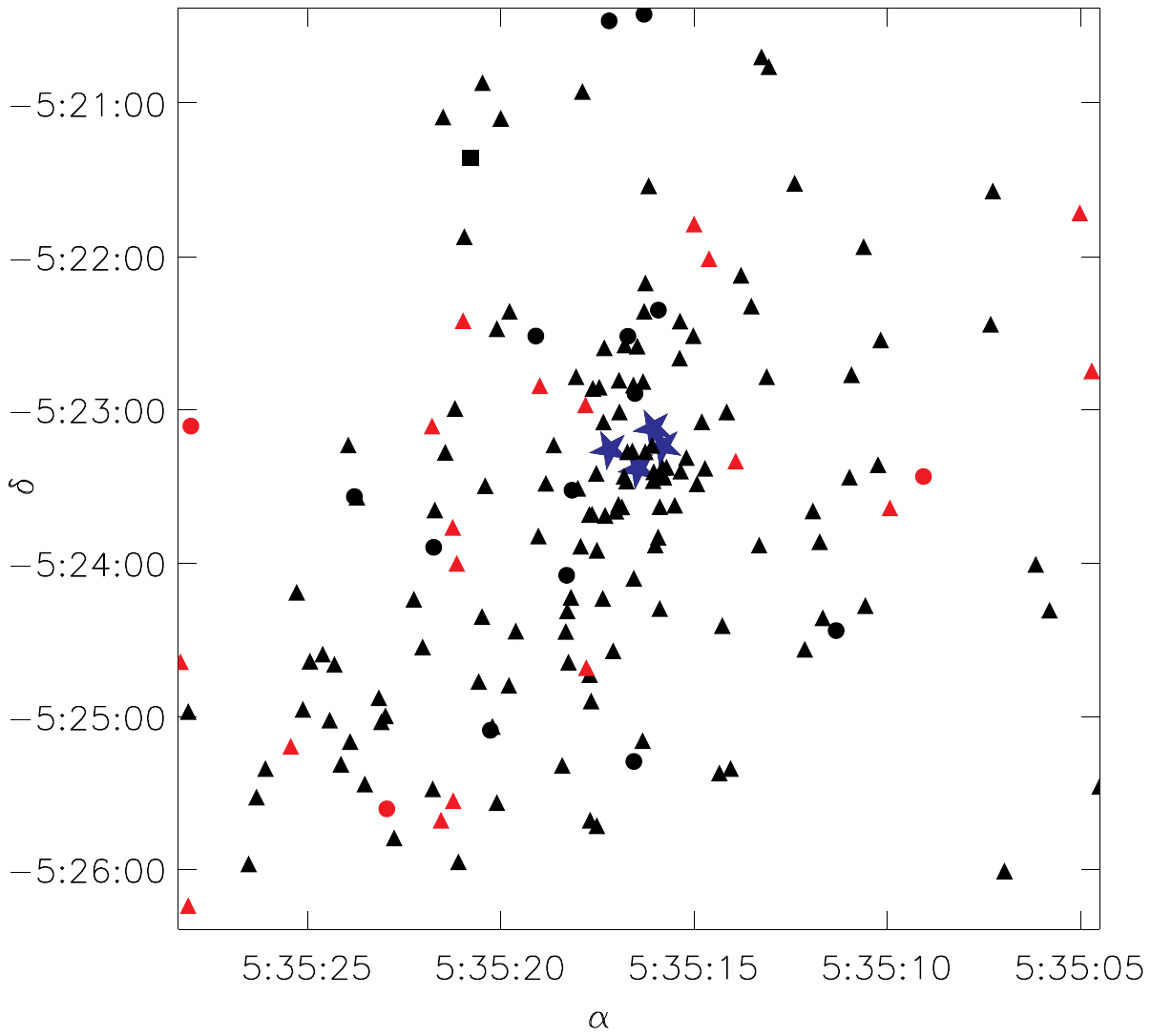}
\caption{Upper Panel: Map of the circumstellar disks and other
extended objects detected in the HST/ACS Treasury Program images.
Right ascension and declination are J2000. The triangles represent
the externally ionized protoplanetary disks, the circles represent
the disks seen only in silhouette, the squares represent the
reflection nebulae with no external ionized gas emission, the
asterisks the sources showing jet emission with neither external
ionized gas emission nor silhouette disk, while the crosses are
the elongated objects described in \S 7. The black objects are the
disks already detected before the observations described by this
paper, while the red color is associated to the new discovered
objects. The dashed lines delimit a $6 \times 6$ arcmin region
centered around $\theta^1$ Ori C, the brightest star of the
Trapezium Cluster. Lower Panel: Expansion of the inner region
delimited by the dashed lines in the upper panel. The four blue
stars are the brightest Trapezium Cluster stars.}\label{map}
\end{figure*}

\begin{figure}
\includegraphics[scale=0.8]{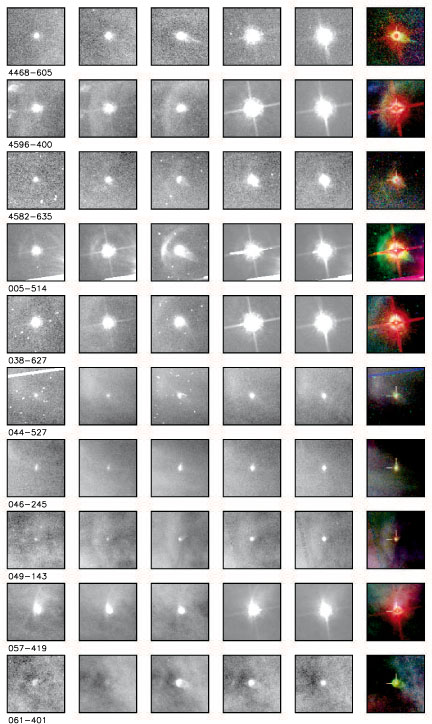}\caption{HST/ACS images
of Orion proplyds (1/18).}{\label{fig_first_ppd}}
\end{figure}

In Figure \ref{map} we show a map with all the circumstellar disks
and the extended objects found in the HST/ACS images. The 63 newly
discovered objects are shown in red. It is remarkable that new
circumstellar disks have been discovered also in the well explored
inner part of M42. Namely, 18 new proplyds and 3 disks seen only
in silhouette have been found in a $6\times6$ arcmin region around
$\theta^1$Ori-C. This demonstrates how important it is to search
for these objects with a multi-wavelength strategy. In several
cases only the presence of a star in our reddest filter images
(F850LP) allows to unambigously recognize the presence a of
circumstellar disks or a reflection nebula too faint to be
detected in the H$\alpha$ filter.

It is evident from Figure~\ref{map} that almost all the disks seen
only in silhouette (the circles in the figure) and  reflection
nebulae (the squares) have been observed in the outskirts of the
Orion Nebula. This for two reasons: 1)\ in the outer regions the
ultraviolet photon flux is low because of the larger distance from
the O- and B-type stars of the Trapezium Cluster (the few
silhouette disks seen in the inner part of the Nebula most
probably lie in the foreground); 2)\ in regions of low nebular
background, it is easier to spot the presence of a disk through
the scattering from the polar regions than by direct imaging of
the dark silhouette against the background. This is how
circumstellar disks are commonly imaged in Tau associations (e.g.
\citet{Kore02}). Conversely, most of the circumstellar disks
associated with emission of externally ionized gas are observed
close to the Trapezium stars. Some them are detected also in the
outer regions, indicating that stars other than  $\theta^1$Ori
(the Trapezium) are affecting the structure and evolution of
protoplanetary disks in the ONC.

In Table \ref{objects_tab} we list all the sources sorted
according to their right ascension and declination, derived from
the absolute astrometric solution of our survey. Cross-references
to previous HST surveys follow the Simbad convention, where the
\citet{Odel94} and \citet{Odel96} lists are merged together and
labelled under the ``OW'' prefix (in this catalog we included also
the four circumstellar disks observed by \citet{Odel01}), whereas
the \citet{Ball00} and \citet{Smit05} lists are merged with the
``BOM'' prefix. We also list the corresponding entry in the
\citet{Pros94} catalog, in the optical survey of \citet{Jone88},
in two near-infrared sources catalogs (\citet{AliD95}; 2 Micron
All-Sky Survey, \citet{Cutr03}) and in the x-ray source catalog of
the Chandra Orion Ultradeep Project (COUP, \citet{Getm05}). The
last 2 columns of Table \ref{objects_tab} report the main
characteristics of the objects derivable from the images. In
particular column (11) defines the type of each object (either
ionized disk seen in emission or dark disk seen only in silhouette
or reflection nebulae with no external ionized gas emission),
while column (12) points out the presence of jets, reflection
nebulae, binary stellar systems, nearly edge-on or face-on
circumstellar disks.

For the object name we used the coordinate-based nomenclature of
\citet{Odel94}: objects with coordinates $\alpha$ = 5:35:AB.C,
$\delta$ = $-5$:2X:YZ are labelled ABC-XYZ. If the right ascension is
5:34:AB.C, then a 4 is added, i.e. 4ABC at the beginning of the
RA\ group. Similarly, if the declination is $-5$:1X:YZ, it becomes
1XYZ. This coordinate-based method is affected by astrometric
errors, as better measures, or just measures at different
wavelengths, may  require a change of name. This is the approach
followed by \citet{Ball00}, who renamed a few sources
originally labelled by the O'Dell team on the basis of their
improved astrometry. Unfortunately, this generates ambiguity and
is a potential source of error when data are retrieved from archives.
For this reason, we decided to maintain the nomenclature of the
objects given in their discovery papers, i.e., in the case of
different names in the OW and BOM catalogues, we used the OW name.
Our coordinates thus take the lowest priority, and were used only for the new objects
discovered by the HST Treasury Program  to give them a name.

Among the 235 circumstellar disks and other extended objects
presented in this paper, 118 have been observed by \citet{AliD95}
in the near infrared; only 49 are listed in the 2MASS catalog
\citep{Cutr03}. The COUP survey shows 137 objects, i.e., 58\% of
all the objects. The COUP fraction rises to 63\% if we do not
consider the extended objects described in \S 7. The high fraction
of circumstellar disks revealed in x-rays is particularly
interesting since, even if the x-ray luminosities are in general
relatively small, this high energy radiation effectively
penetrates deeper through the disks, ionizing otherwise neutral
molecular gases and even melting solid particles. Together with
the ionizing flux from the brightest cluster members, x-rays from
low mass  star may thus have profound effects on their associated
circumstellar disks and therefore on planet formation
\citep{Feig07}.

\section{New Proplyds}

We detected 29 previously unknown proplyds, whose images have been
reported together with all the other Orion proplyds in Figures
\ref{fig_first_ppd}-\ref{fig_last_ppd}. In Table \ref{objects_tab}
they can be recognized as those ionized disks seen in emission
(flag ``i'' in Column (11)) that have not been observed in the OW
and BOM catalogs (i.e. with no designations in Columns (4) and
(5), except for the proplyd 280-931, observed by \citet{Ball01}
not included in the BOM catalog in Simbad).

Among these 29 new proplyds, 3 show evidence of emission from jets
mainly in the H$\alpha$ filter, and all of them are located in the
outskirts of the Orion Nebula (4468-605, 099-339, 351-349). Also,
other 5 objects in the M42 outer regions show evidence of jet
emission in $H\alpha$ (4364-146, 4466-324, 006-439, 078-3658,
353-130, see Figure \ref{fig_jet}). This is probably due to the
low level of background nebular emission in those regions that
makes the faint jets easier to be detected than in the inner
region of M42.

Compared with many of the previously known proplyds with bright
cusps observed in the M42 core, these new objects are fainter. For
the proplyds located in the outer regions of the Orion Nebula,
this lack of ionized gas is due to the distance from the ionizing
sources located in the M42 core. The lack of bright cusps in
proplyds located in the inner region can be explained by several
factors: their physical distance from the M42 core may be larger
than the projected one due to the position of these objects with
respect to the line of sight, these disks may have a smaller
amount of mass compared with the brighter proplyds, the
photo-evaporation processes in act in the disks surface may be in
an early or late phase, so that the ionization front is not much
developed.

In the following we briefly describe two of the new proplyds with
bright ionization fronts, located $\sim 10'$-far from the ONC
core.

\textbf{064-3335}: (Fig. 3, row 1) This proplyd is located $\sim
10'$ south of the ONC core. Its ionization cusp has a diameter of
roughly $3.5''$ with a P. A. of $\sim 300^{\circ}$. Other than the
bright cusp, observed in all the 5 ACS filters, two filaments
extending for $\sim 800$ AU from the center of the proplyd are
visible mainly in the H$\alpha$ filter, almost along the cusp axis
direction.

\textbf{066-3251}: (Fig. 3, row 2) This proplyd is located $\sim
10'$ south of the ONC core and it is very close to 064-3335 (the
distance between the 2 proplyds is $43''$). The ionization cusp,
oriented with a P. A. $\sim 320^{\circ}$, has a diameter of
roughly $3''$. Respect to 066-3251 this bright star is located
$225''$ on a direction P. A. $\sim 315^{\circ}$. In the southern
side of the object a long outflow extends for about $30''$,
corresponding to $\sim 600$ AU.

\section{New Silhouette Disks and Reflection Nebulae}

In this paragraph we provide a short description of the 7 dark
silhouette disks and the 6 reflection nebulae discovered by the
new HST/ACS images.

\textbf{090-326}: (Fig. 20, row 2) This silhouette disk is located
$\sim 10'$ southeast of ONC core. It has a P. A. of roughly
$50^{\circ}$, major and minor axes of about $\sim 0.3''$ and
$0.15''$ in the H$\alpha$ filter (approximately 120 $\times$ 60
AU) respectively, from which an inclination angle of $\sim
60^{\circ}$ can be derived assuming a circular thin disk. However
since the central star is obscured in the F435W, F555W, and
H$\alpha$ filters, the disk may have larger inclination angle and
thickness. In the bluer filters a faint emission is detected in
the southeast side all around the disk. The emission from the disk
edges can be due either to reflection nebular light, or to a very
mild level of ionization of the disk surface, in a region of the
Orion Nebula where the UV flux from the O- and B-spectral type
stars is unable to support a fully developed proplyd.

\textbf{230-536}: (Fig. 21, row 8) This small silhouette disk
(approximately $0.5''$ $\times$ $0.25''$ in the F435W filter,
correspondingly to about $200$ $\times$ $100$ AU), located $\sim
5'$ southeast of ONC core has an inclination angle of roughly
$\sim 60^{\circ}$ as derived from the apparent axes ratio. This
justifies the detection of the red pre-sequence-star in the
H$\alpha$ filter. Its P.A. is about $\sim 160^{\circ}$.

\textbf{280-1720}: (Fig. 22, row 1) This silhouette disk, located
$\sim 8'$ northwest of the ONC core is seen nearly face-on in the
F435W and H$\alpha$ filters . The size is approximately $0.75''$
$\times$ $0.75''$ in the H$\alpha$ filter, corresponding to about
$300$ $\times$ $300$ AU.

\textbf{281-306}: (Fig. 22, row 2) This small silhouette disk,
located $\sim 12'$ southeast of the Trapezium Cluster is visible
face-on only in the H$\alpha$ filter. The diameter of the disk is
about $\sim 0.4''$ or about $160$ AU.

\textbf{332-405}: (Fig. 22, row 5) This silhouette disk, located
$\sim 5'$ east of the ONC core is seen in absorbtion in the F435W,
F555W and H$\alpha$ filters. The major and minor axes are
approximately $0.75''$ $\times$ $0.25''$, corresponding to about
$300$ $\times$ $100$ AU in the F555W filter, and implying an
inclination angle of about $70^{\circ}$. The disk P. A. is $\sim
120^{\circ}$.

\textbf{346-1553}: (Fig. 22, row 6) This silhouette disk, located
in the M43 region, $\sim 12'$ northeast of the ONC core, is seen
in absorbtion only in the broadband filters F435W and F555W. The
disk appears to be face-on with a diameter of $\sim 0.5''$ in the
F435W, or $\sim 200$ AU at the distance of Orion Nebula.

\textbf{473-245}: (Fig. 22, row 8) This spectacular silhouette
disk with reflection nebula is located $\sim 10'$ east of the ONC
core. The disk is seen nearly edge-on (the two sides of the
bipolar emission appear to be very symmetric), flared, with a P.A.
of about $\sim 60^{\circ}$ and a major axis in the H$\alpha$ of
roughly $0.75''$, corresponding to a physical diameter of 300 AU.

\textbf{4538-311}: (Fig. 23, row 1) The disk, located $\sim 6'$
east of the ONC core, appears as an equatorial dark lane at the
midplane of bipolar nebula. The emission is nearly symmetric, with
the northeast side
 slightly wider and brighter than the southwest one, suggesting that
it originates from the surface of a disk seen almost edge-on with
the northeast face tilted toward us. The nebula is detected only
in the F775W and F850LP filters, with some emission from the
northwest side  visible also in the F555W close to the noise floor
of our image. This indicates that the pre-main-sequence star
hidden by the disk is very red. The P. A. of the disk is
approximately $150^{\circ}$.

\textbf{016-149}: (Fig. 23, row 2) This object, located $\sim 3'$
northeast of the ONC core, appears only in the F775W and F850LP
filters as a bipolar nebula. In this case the morphology is highly
asymmetric, with the southwest side much brighter and more
extended than the northeastern one. The asymmetry, together with
the low contrast of the equatorial dust lane, suggests that the
disk is seen with low inclination angle, i.e. far from being
face-on. However, the fact that the northeastern lobe appears as a
point-like source brighter than the more extended southwestern one
suggests that the former source may rather be a red star, whose
radiation scattered by circumstellar matter is seen as the
southwestern lobe.

\textbf{046-3838}: (Fig. 23, row 3) This source, $\sim 15'$ south
of the ONC core, shows an extended region (diameter $\sim 5''$)
with bright emission especially in the F775W and F850LP filters,
and a dark tail in the direction P. A. $\sim 350^{\circ}$. Since
046-3838 is rather weak in H$\alpha$ compared to the fluxes
observed in the F555W, F775W and F850LP, this source is most
probably a reflection nebula with a red central star.

\textbf{051-3541}: (Fig. 23, row 4) Located $\sim 15'$ south of
the ONC core, this  bipolar source appears highly symmetric and
relatively bright. It is detected mainly in the F775W and F850LP
filters with a conspicuous equatorial dust lane at P. A. of about
$100^{\circ}$, suggesting the presence of an almost edge-on disk
around a red pre-main-sequence star. The lobes appear nearly round
and more extended than those of 4538-311, for instance, suggesting
either a very strong disk flaring or the presence of circumstellar
material at the disk polar regions.

\textbf{193-1659}: (Fig. 23, row 6) This well developed bipolar
nebula is located $\sim 7'$ north of the ONC core . The
asymmetrical brightness of the two lobes (seen only in the F775W
and F850LP filters) and the detection of a red pre-main-sequence
star away from the center of the bipolar nebula suggest that
either the circumstellar disk that blocks the star light or the
circumstellar matter that reflect it is non symmetric. The P. A.
of the disk is approximately $100^{\circ}$.

\textbf{294-757}: (Fig. 23, row 8) This red source, observed only
in the F775W and F850LP $\sim 7'$ southeast of the ONC core, is a
bipolar nebula with equatorial dark lane clearly detected in the
F775W image. This suggests the presence of a nearly edge-on disk
with P. A. $\sim 70^{\circ}$ (southeastern side facing toward us).

\section{Candidate background galaxies}

In Figures \ref{fig_first_gal}-\ref{fig_last_gal}, 16 red and
elongated objects are shown. These sources are listed in Table
\ref{other_objects_tab}. They are all visible only in the F775W
and F850LP filters and in the outer regions of the Orion Nebula
(see Figure \ref{map}). These two facts suggest that these objects
might be background galaxies seen through sparse regions of the
reddening Orion Molecular Cloud OMC-1. An alternative plausible
interpretation is that of reflection nebulae turned on by red
pre-main-sequence stars (two similar objects in which nearly
edge-on reflecting disks are visible are 294-757 and 016-149). In
this hypothesis, their location in the Orion Nebula outskirts, far
from the ionizing O- and B-type stars in the Trapezium Cluster, is
consistent with the non-detection of light from ionized plasma,
that would have been observed in the H$\alpha$ filter as well. To
understand the nature of these objects spectra are needed.

\section{Summary}

In this paper we have shown the HST/ACS images of the 178
proplyds, 28 disks seen only in silhouette, 8 reflection nebulae
without external ionized plasma, 5 jets without neither external
ionized plasma nor silhouette disk and 16 other extended objects
observed by the HST Treasury Program on the Orion Nebula Cluster.
For every object we have reported all the images taken through the
5 photometric filters used by the HST Treasury Program (F435W,
F555W, F658N, F775W, F850LP).

The fact that most of these objects are associated to X-rays
sources observed by the Chandra Orion Ultradeep Project is
particularly interesting, since high energy photons could play an
important role in the star and planet formation processes.

Among all the objects reported, 63 have been discovered by these
images: 29 proplyds, 7 silhouette disks, 6 reflection nebulae with
no external ionized plasma, 5 jets with no external ionized plasma
or silhouette disk, and 16 other elongated object.

Searching in the literature we found that 4 objects previously
reported as circumstellar disks have not been detected by HST/ACS
images either because hidden by the saturation bleeding trails of
a close bright star or because located out of HST/ACS Treasury
Program field of view. For other 30 sources previously reported as
extended objects HST/ACS images reveal no circumstellar emission
around them.

A brief description of all the newly discovered proplyds, disks
seen only in silhouette and reflection nebulae with no external
ionized plasma has been carried out in \S 5 and \S 6.

Finally, we have discussed possible interpretations for the nature
of the 16 extended objects. Because of their location far from the
Trapezium Cluster ($\gtrsim 10'$) and because of their red color,
they are probably background galaxies reddened by the Orion
Molecular Cloud OMC-1, but the alternative hypothesis of
reflection nebulae turned on by red pre-main-sequence stars cannot
be ruled out by our observations only.
\\
\\
We wish to thank the referee, William Henney, for his useful
comments that greatly improved the manuscript. Support for program
10246 was provided by NASA through a grant from the Space
Telescope Science Institute, which is operated by the Association
of Universities for Research in Astronomy, Inc., under NASA
contract NAS 5-26555. The work of LR at STScI was done under the
auspices of the STScI Summer Student Program.




\clearpage




\clearpage

\clearpage

\clearpage

\begin{figure}
\epsscale{.80}
\includegraphics[scale=0.8]{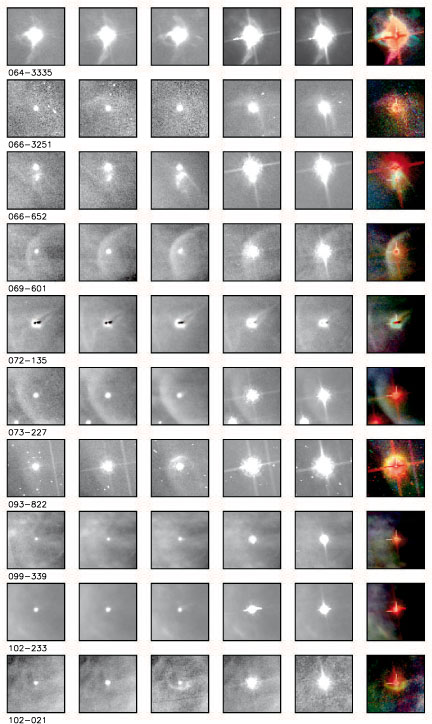} \caption{HST/ACS images
of Orion proplyds (2/18).}
\end{figure}
\clearpage

\begin{figure}
\epsscale{.80}
\includegraphics[scale=0.8]{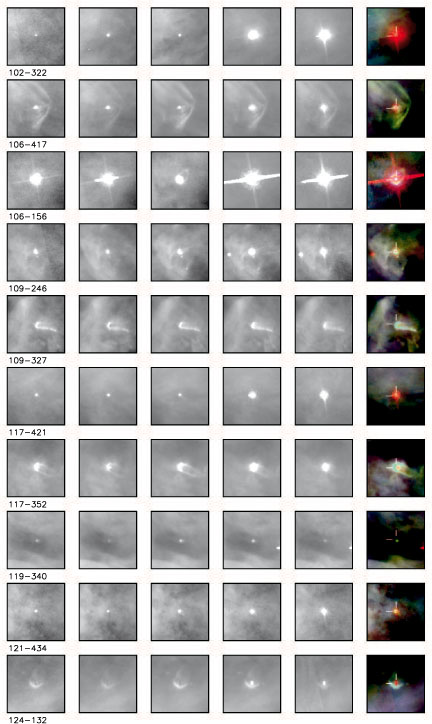} \caption{HST/ACS images
of Orion proplyds (3/18).}
\end{figure}
\clearpage

\begin{figure}
\epsscale{.80}
\includegraphics[scale=0.8]{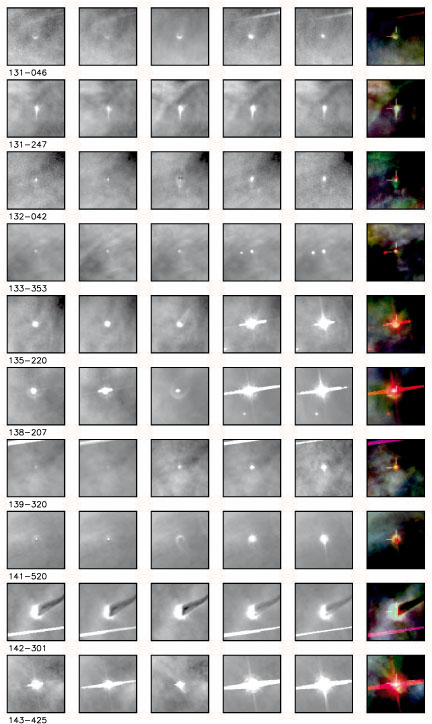} \caption{HST/ACS images
of Orion proplyds (4/18).}
\end{figure}
\clearpage

\begin{figure}
\epsscale{.80}
\includegraphics[scale=0.8]{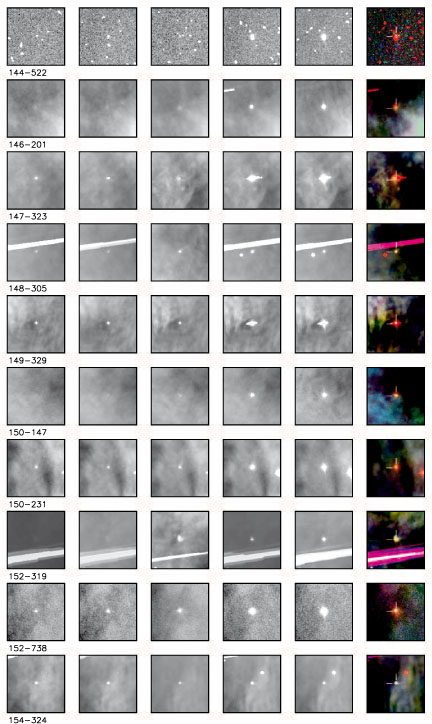} \caption{HST/ACS images
of Orion proplyds (5/18).}
\end{figure}
\clearpage

\begin{figure}
\epsscale{.80}
\includegraphics[scale=0.8]{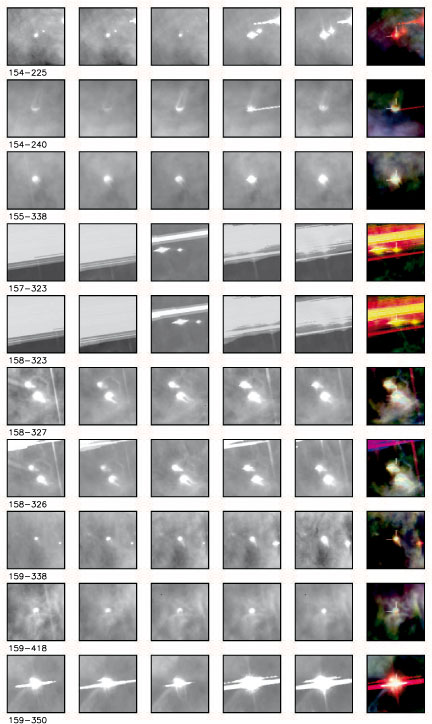} \caption{HST/ACS
images of Orion proplyds (6/18).}
\end{figure}
\clearpage

\begin{figure}
\epsscale{.80}
\includegraphics[scale=0.8]{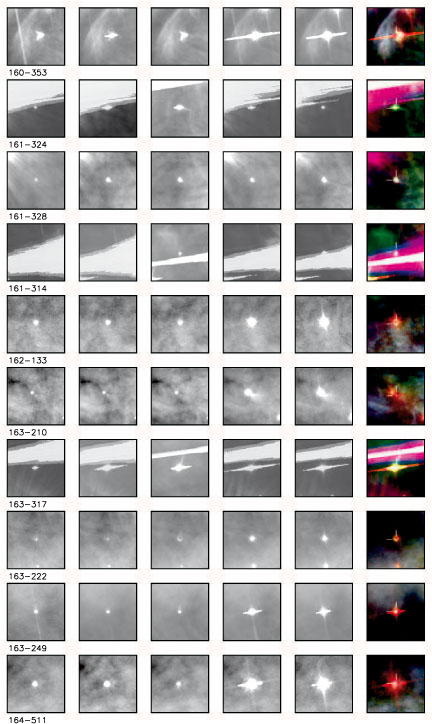} \caption{HST/ACS images
of Orion proplyds (7/18).}
\end{figure}
\clearpage

\begin{figure}
\epsscale{.80}
\includegraphics[scale=0.8]{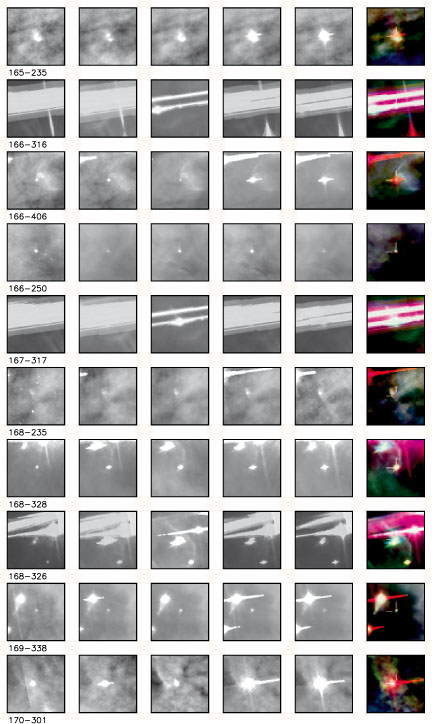} \caption{HST/ACS images
of Orion proplyds (8/18).}
\end{figure}
\clearpage

\begin{figure}
\epsscale{.80}
\includegraphics[scale=0.8]{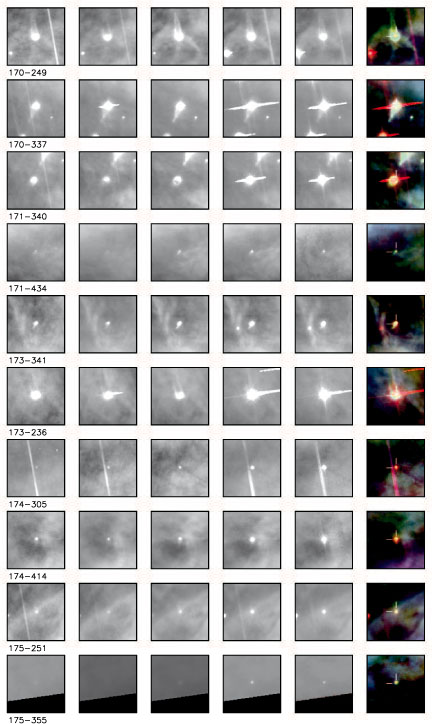} \caption{HST/ACS images
of Orion proplyds (9/18).}
\end{figure}
\clearpage

\begin{figure}
\epsscale{.80}
\includegraphics[scale=0.8]{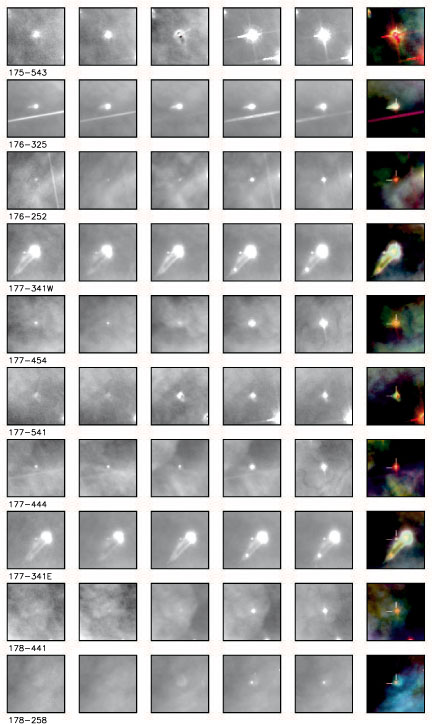} \caption{HST/ACS images
of Orion proplyds (10/18).}
\end{figure}
\clearpage

\begin{figure}
\epsscale{.80}
\includegraphics[scale=0.8]{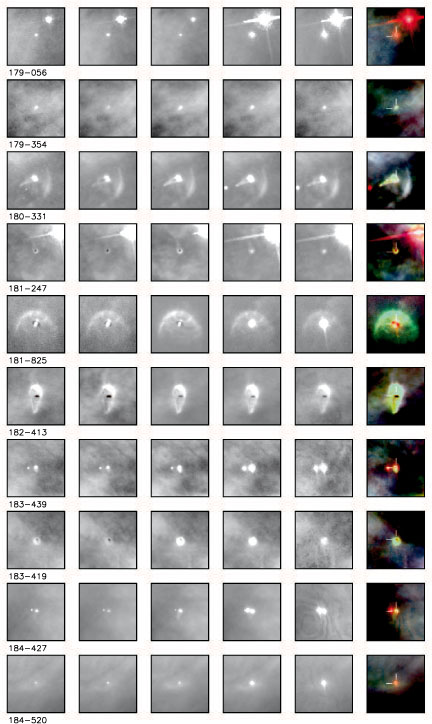} \caption{HST/ACS images
of Orion proplyds (11/18).}
\end{figure}
\clearpage

\begin{figure}
\epsscale{.80}
\includegraphics[scale=0.8]{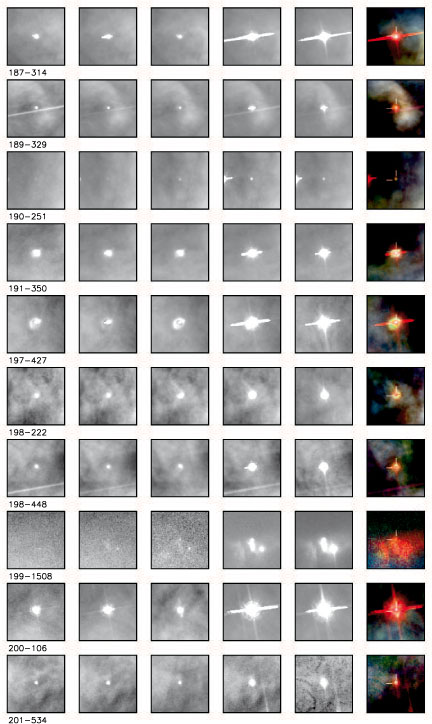} \caption{HST/ACS images
of Orion proplyds (12/18).}
\end{figure}
\clearpage

\begin{figure}
\epsscale{.80}
\includegraphics[scale=0.8]{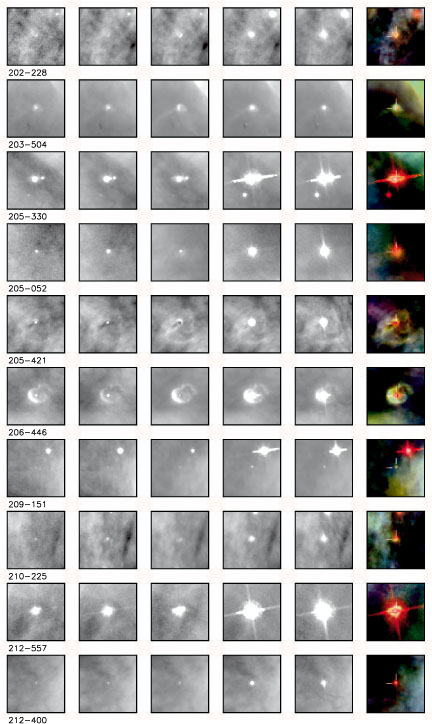} \caption{HST/ACS images
of Orion proplyds (13/18).}
\end{figure}
\clearpage

\begin{figure}
\epsscale{.80}
\includegraphics[scale=0.8]{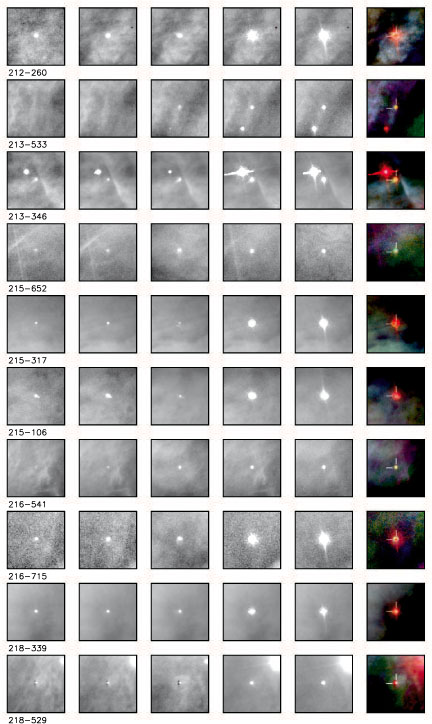} \caption{HST/ACS images
of Orion proplyds (14/18).}
\end{figure}
\clearpage

\begin{figure}
\epsscale{.80}
\includegraphics[scale=0.8]{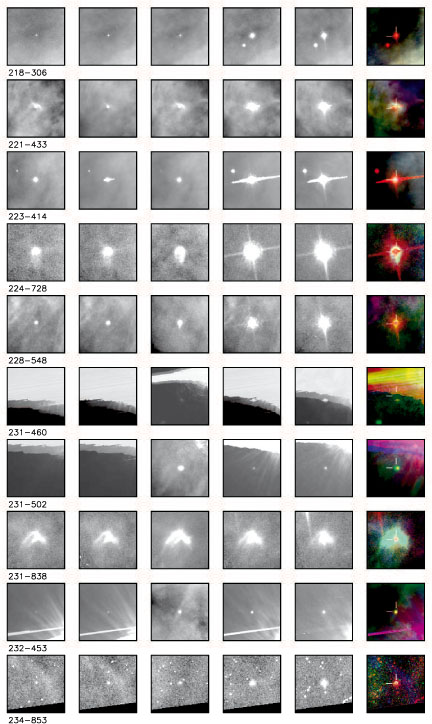} \caption{HST/ACS images
of Orion proplyds (15/18).}
\end{figure}
\clearpage

\begin{figure}
\epsscale{.80}
\includegraphics[scale=0.8]{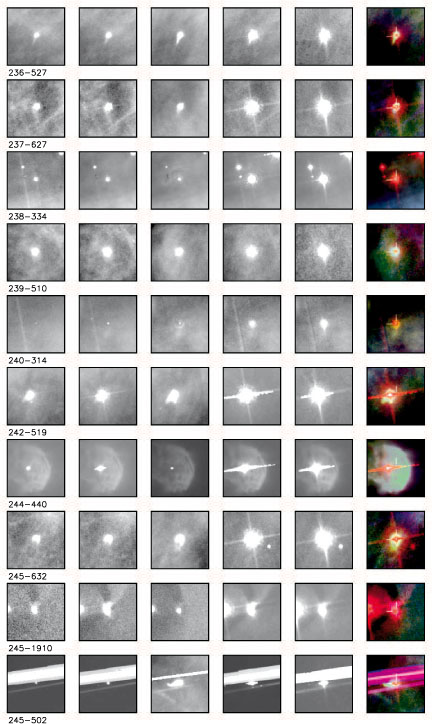} \caption{HST/ACS images
of Orion proplyds (16/18).}
\end{figure}
\clearpage

\begin{figure}
\epsscale{.80}
\includegraphics[scale=0.8]{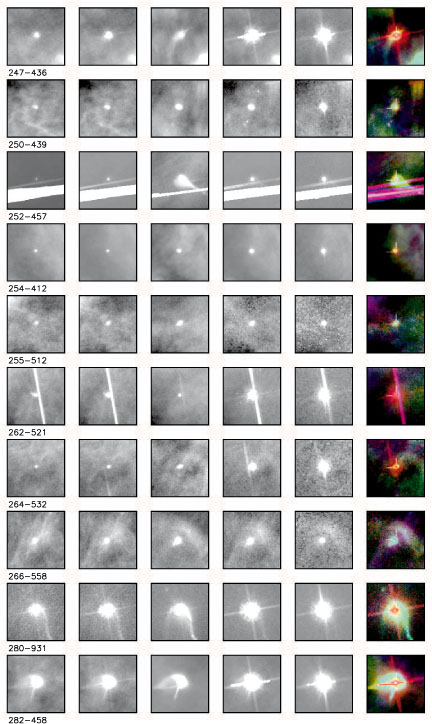} \caption{HST/ACS images
of Orion proplyds (17/18).}
\end{figure}
\clearpage

\begin{figure}
\epsscale{.80}
\includegraphics[scale=0.8]{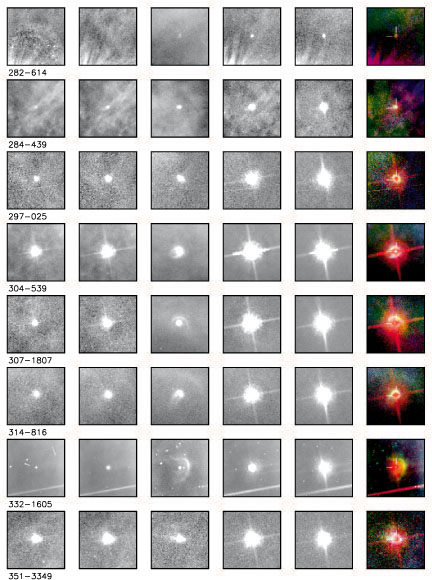} \caption{HST/ACS images
of Orion proplyds (18/18).}{\label{fig_last_ppd}}
\end{figure}
\clearpage

\begin{figure}
\epsscale{.80}
\includegraphics[scale=0.8]{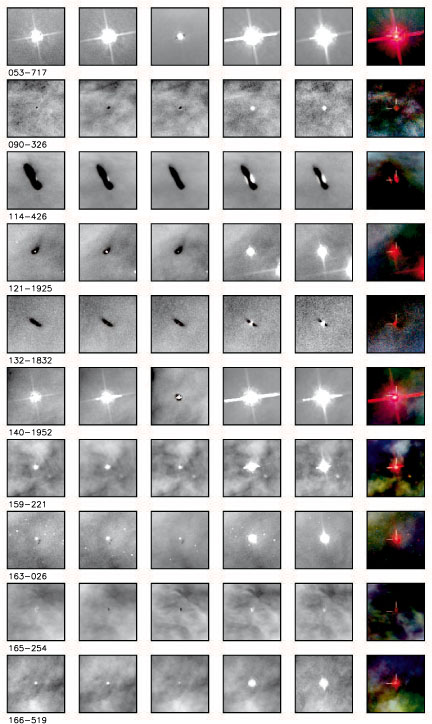} \caption{HST/ACS images
of Orion disks seen only in silhouette
(1/3).}{\label{fig_first_sd}}
\end{figure}
\clearpage

\begin{figure}
\epsscale{.80}
\includegraphics[scale=0.8]{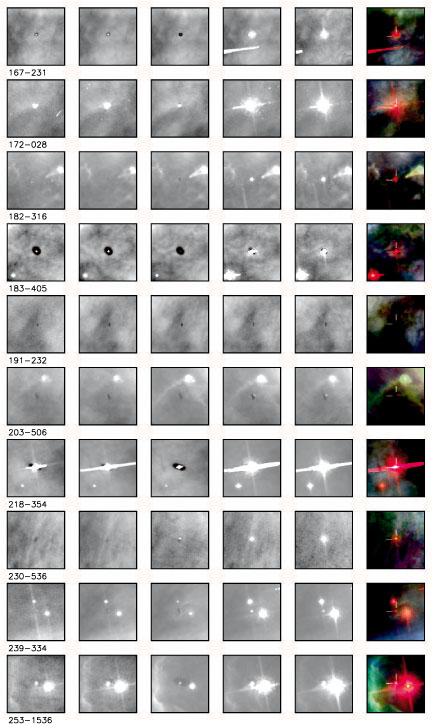} \caption{HST/ACS images
of Orion disks seen only in silhouette (2/3).}
\end{figure}
\clearpage

\begin{figure}
\epsscale{.80}
\includegraphics[scale=0.8]{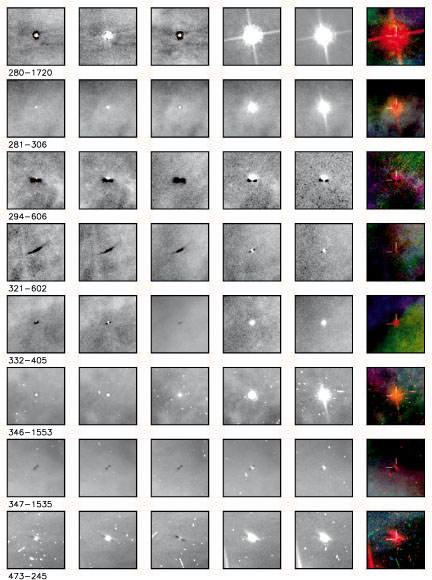} \caption{HST/ACS images
of Orion disks seen only in silhouette
(3/3).}{\label{fig_last_sd}}
\end{figure}
\clearpage

\begin{figure}
\epsscale{.80}
\includegraphics[scale=0.8]{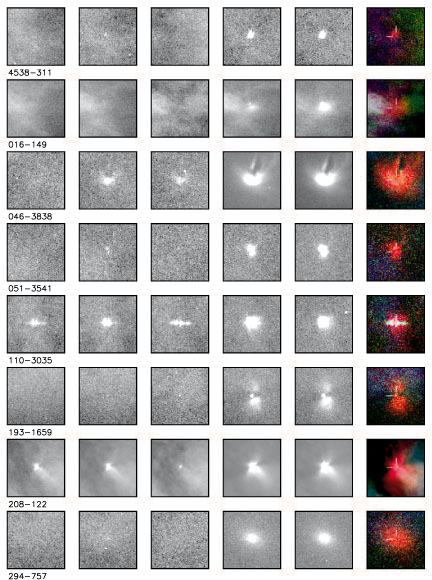} \caption{HST/ACS images
of Orion reflection nebulae with no external ionized
gas.}{\label{fig_rn}}
\end{figure}
\clearpage

\begin{figure}
\epsscale{.80}
\includegraphics[scale=0.8]{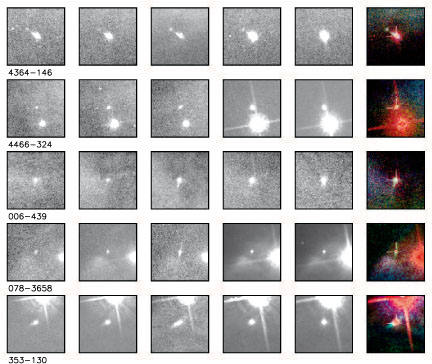} \caption{HST/ACS images
of Orion sources with jet emission with no evidence of neither
external ionized gas emission nor silhouette
disk.}{\label{fig_jet}}
\end{figure}
\clearpage

\begin{figure}
\epsscale{.80}
\includegraphics[scale=0.8]{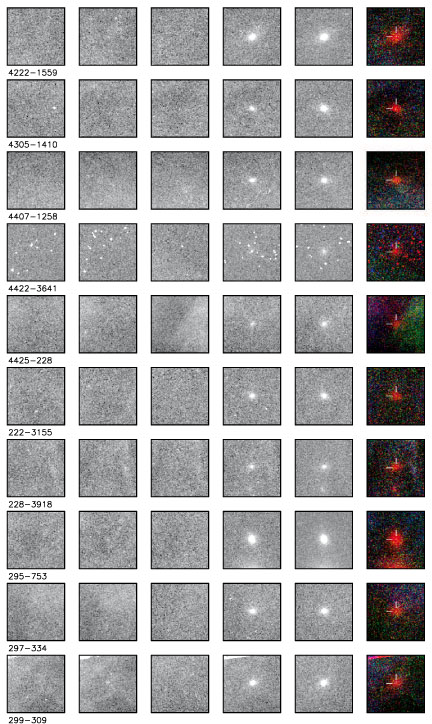} \caption{HST/ACS images
of galaxy candidates (1/2).}{\label{fig_first_gal}}
\end{figure}
\clearpage

\begin{figure}
\epsscale{.80}
\includegraphics[scale=0.8]{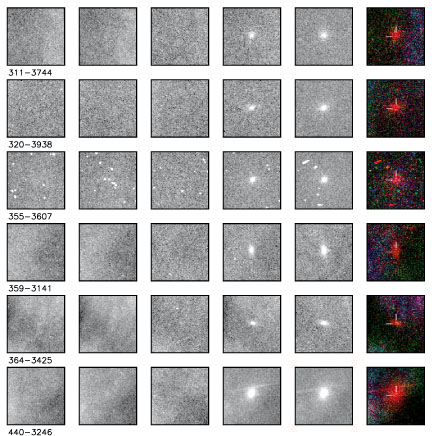} \caption{HST/ACS images
of galaxy candidates (2/2).}{\label{fig_last_gal}}
\end{figure}

\clearpage

\begin{figure}
\epsscale{.80}
\includegraphics[scale=0.8]{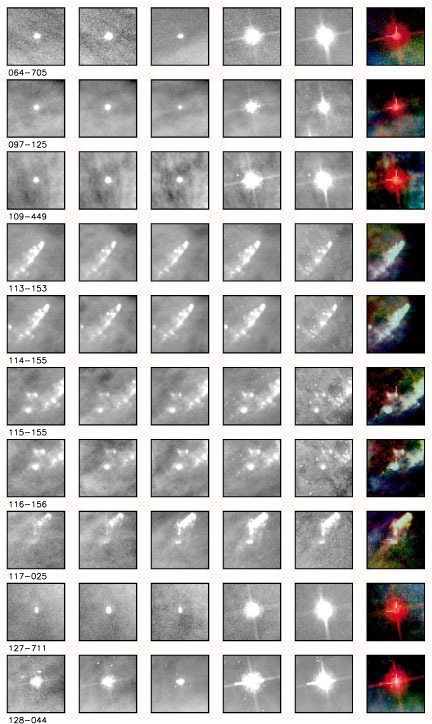} \caption{HST/ACS
images of the objects previously identified as non stellar
excluded from our catalogue of circumstellar disks
(1/4).}{\label{fig_first_other}}
\end{figure}
\clearpage

\begin{figure}
\epsscale{.80}
\includegraphics[scale=0.8]{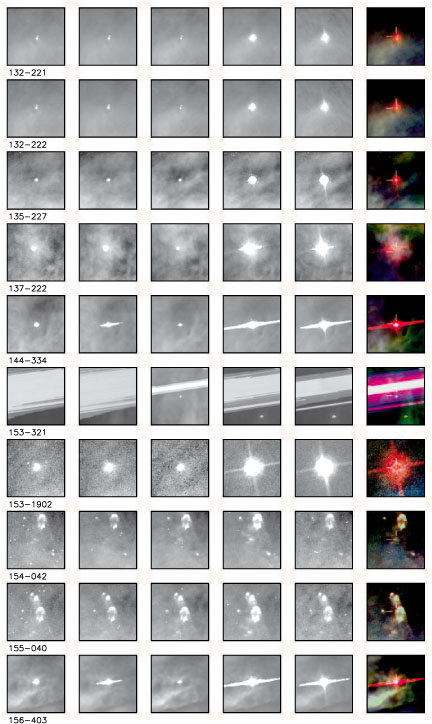} \caption{HST/ACS
images of the objects previously identified as non stellar
excluded from our catalogue of circumstellar disks (2/4).}
\end{figure}
\clearpage

\begin{figure}
\epsscale{.80}
\includegraphics[scale=0.8]{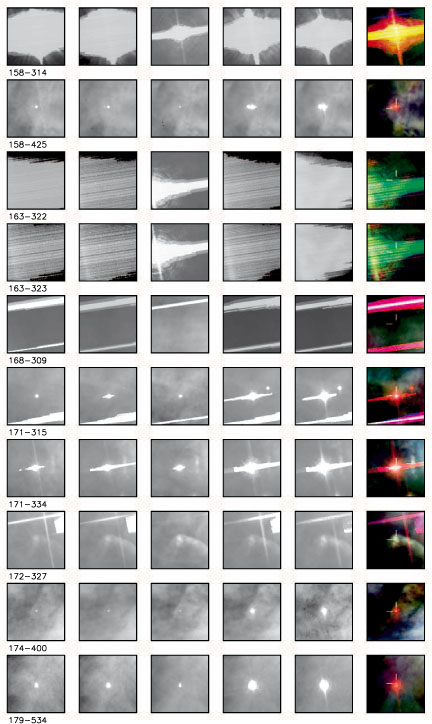} \caption{HST/ACS
images of the objects previously identified as non stellar
excluded from our catalogue of circumstellar disks (3/4).}
\end{figure}
\clearpage

\begin{figure}
\epsscale{.80}
\includegraphics[scale=0.8]{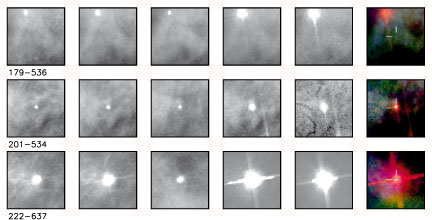} \caption{HST/ACS
images of the objects previously identified as non stellar
excluded from our catalogue of circumstellar disks
(4/4).}{\label{fig_last_other}}
\end{figure}

\clearpage

\begin{deluxetable}{lcc}
\tablecaption{ACS/WFC Photometric Filters\label{Tab:Filters}}
\tablewidth{0pt} \tablehead{\colhead{Filter} & \colhead{Ground
Equivalent} & \colhead{Integration time (s)} } \startdata
F435W & Johnson $B$ & 420 \\
F555W & Johnson $V$ & 385 \\
F658N & H$\alpha$+[N II] $\lambda$6583 & 340 \\
F775W & Cousin $I_{\rm C}$ & 385 \\
F850LP & $z$-band & 385 \\
\enddata\
\end{deluxetable}

\begin{deluxetable}{lccc}
\tablecaption{ACS/WFC Photometric Zero
Points\label{Tab:Zero_Points}} \tablewidth{0pt}
\tablehead{\colhead{Filter} & \colhead{VEGAMAG} & \colhead{ABMAG}
& \colhead{STMAG} } \startdata
F435W & 25.779 & 25.673 & 25.157 \\
F555W & 25.724 & 25.718 & 25.672 \\
F658N & 22.365 & 22.747 & 23.148 \\
F775W & 25.256 & 25.654 & 26.393 \\
F850LP & 24.326 & 24.862 & 25.954 \\
\enddata\
\end{deluxetable}

\clearpage
\begin{deluxetable}{lccccccccccc}
\tabletypesize{\scriptsize} \rotate
\renewcommand{\arraystretch}{0.6}
\tablecaption{Circumstellar disks from the HST/ACS Treasury
Program.\label{objects_tab}} \tablewidth{0pt} \tablehead{
\colhead{Object} & \colhead{RA\tablenotemark{a}} &
\colhead{DEC\tablenotemark{a}} & \colhead{OW\tablenotemark{b}} &
\colhead{BOM\tablenotemark{b}} & \colhead{JW\tablenotemark{b}}&
\colhead{P\tablenotemark{b}}& \colhead{AD\tablenotemark{b}}&
\colhead{2MASS\tablenotemark{b}}& \colhead{COUP\tablenotemark{b}}&
\colhead{Type\tablenotemark{c}} & \colhead{Note\tablenotemark{d}}}
\startdata

4364-146    &    5:34:36.44     &    -5:21:45.95    &   -   &   -   &   -   &   -   &   -   &   J05343646-0521458   &   -   &   j   &   J     \\
4466-324    &    5:34:46.59     &    -5:23:24.19    &   -   &   -   &   83  &   -   &   987 &   J05344656-0523256   &   29  &   j    &  J,B   \\
4468-605    &    5:34:46.76     &    -5:26:04.79    &   -   &   -   &   86  &   -   &   95  &   -   &   -   &   i   &   J     \\
4538-311    &    5:34:53.79     &    -5:23:10.73    &   -   &   -   &   -   &   -   &   -   &   -   &   -   &   rn  &         \\
4596-400    &   5:43:59.56  &   -5:24:00.19 &   4596-400    &   000-400 &   191 &   -   &   1026    &   J05345955-0524002   &   137 &   i   &         \\
4582-635    &    5:34:58.16     &    -5:26:35.13    &   -   &   -   &   -   &   -   &   -   &   J05345816-0526350   &   -   &   i   &         \\
005-514 &    5:35:00.47     &    -5:25:14.34    &   005-514 &   005-514 &   198 &   -   &   875 &   J05350046-0525143   &   147 &   i   &         \\
006-439 &    5:35:00.58     &    -5:24:38.79    &   -   &   -   &   -   &   -   &   -   &   -   &   -   &   j   &   J     \\
016-149 &    5:35:01.60     &    -5:21:49.35    &   -   &   -   &   -   &   -   &   1063    &   J05350162-0521489   &   165 &   rn  &         \\
038-627 &    5:35:04.19     &    -5:26:27.89    &   038-627 &   -   &   245 &   -   &   859 &   J05350419-0526278   &   212 &   i   &         \\
044-527 &    5:35:04.42     &    -5:25:27.40    &   044-527 &   044-527 &   -   &   -   &   -   &   J05350445-0525264   &   -   &   i   &         \\
046-3838    &    5:35:04.61     &    -5:38:38.00    &   -   &   -   &   -   &   -   &   92  &   J05350461-0538379   &   -   &   rn   &         \\
046-245 &    5:35:04.63     &    -5:22:44.85    &   -   &   -   &   -   &   -   &   -   &   -   &   -   &   i   &         \\
049-143 &    5:35:04.94     &    -5:21:42.99    &   -   &   -   &   -   &   -   &   -   &   -   &   -   &   i   &         \\
051-3541    &    5:35:05.05     &    -5:35:40.84    &   -   &   -   &   -   &   -   &   -   &   J05350505-0535407   &   -   &   rn  &   EO    \\
053-717 &    5:35:05.40     &    -5:27:16.99    &   -   &   053-717 &   268 &   -   &   845 &   J05350540-0527170   &   241 &   d   &         \\
057-419 &    5:35:05.73     &    -5:24:18.55    &   057-419 &   -   &   278 &   -   &   -   &   J05350572-0524184   &   250 &   i   &         \\
061-401 &    5:35:06.09     &    -5:24:00.60    &   061-401 &   -   &   -   &   -   &   -   &   -   &   -   &   i   &         \\
064-3335    &    5:35:06.44     &    -5:33:35.25    &   -   &   -   &   295 &   -   &   2221    &   J05350644-0533351   &   267 &   i   &         \\
066-3251    &    5:35:06.57     &    -5:32:51.49    &   -   &   -   &   -   &   -   &   371 &   J05350656-0532515   &   273 &   i   &         \\
066-652 &    5:35:06.59     &    -5:26:51.99    &   066-652 &   -   &   296 &   -   &   851 &   J05350660-0526509   &   275 &   i   &   B     \\
069-601 &    5:35:06.91     &    -5:26:00.60    &   069-601 &   069-601 &   299 &   -   &   867 &   -   &   279 &   i   &         \\
072-135 &    5:35:07.21     &    -5:21:34.43    &   072-135 &   072-135 &   -   &   -   &   1069    &   -   &   -   &   i   &   EO    \\
073-227 &    5:35:07.27     &    -5:22:26.56    &   073-227 &   073-227 &   300 &   -   &   3012    &   -   &   283 &   i   &         \\
078-3658    &    5:35:07.84     &    -5:36:58.15    &   -   &   -   &   -   &   -   &   -   &   -   &   -   &   j   &   J,B   \\
090-326 &    5:35:09.02     &    -5:23:26.20    &   -   &   -   &   -   &   -   &   -   &   -   &   -   &   d   &         \\
093-822 &    5:35:09.59     &    -5:28:22.92    &   093-822 &   -   &   334 &   -   &   2257    &   J05350959-0528228   &   341 &   i   &         \\
099-339 &    5:35:09.89     &    -5:23:38.50    &   -   &   -   &   -   &   2   &   2654    &   -   &   350 &   i   &   J     \\
102-233 &    5:35:10.13     &    -5:22:32.74    &   102-233 &   -   &   340 &   -   &   3354    &   -   &   358 &   i   &         \\
102-021 &   5:35:10.19  &   -5:20:20.99 &   102-021 &   -   &   339 &   -   &   3357    &   J05351019-0520210   &   362 &   i   &         \\
102-322 &    5:35:10.20     &    -5:23:21.56    &   102-322 &   -   &   341 &   -   &   3148    &   J05351021-0523215   &   365 &   i   &         \\
106-417 &    5:35:10.54     &    -5:24:16.70    &   106-417 &   -   &   349 &   4   &   3291    &   -   &   385 &   i   &         \\
106-156 &    5:35:10.58     &    -5:21:56.24    &   106-156 &   -   &   347 &   -   &   3181    &   J05351058-0521562   &   382 &   i   &         \\
109-246 &    5:35:10.90     &    -5:22:46.36    &   109-246 &   109-247 &   355 &   -   &   3476    &   -   &   403 &   i   &         \\
109-327 &   5:35:10.95  &   -5:23:26.45 &   109-327 &   109-327 &   -   &   -   &   3401    &   -   &   -   &   i   &   J     \\
110-3035    &    5:35:10.98     &    -5:30:35.23    &   -   &   110-3035    &   -   &   -   &   -   &   -   &   -   &   rn  &   J     \\
114-426 &    5:35:11.30     &    -5:24:26.50    &   114-426 &   114-426 &   -   &   -   &   2578    &   -   &   419 &   d   &   EO,RN     \\
117-421 &    5:35:11.65     &    -5:24:21.50    &   117-421 &   -   &   366 &   6   &   2588    &   -   &   434 &   i   &         \\
117-352 &    5:35:11.73     &    -5:23:51.70    &   -   &   117-352 &   368 &   7   &   2641    &   -   &   443 &   i   &         \\
119-340 &   5:35:11.90  &   -5:23:39.70 &   119-340 &   -   &   -   &   -   &   -   &   -   &   -   &   i   &         \\
121-1925    &    5:35:12.09     &    -5:19:24.80    &   121-1925    &   121-1925    &   374 &   -   &   -   &   -   &   460 &   d   &         \\
121-434 &    5:35:12.12     &    -5:24:33.80    &   121-434 &   -   &   376 &   -   &   2569    &   -   &   465 &   i   &         \\
124-132 &    5:35:12.38     &    -5:21:31.39    &   124-132 &   124-132 &   -   &   17  &   2840    &   -   &   476 &   i   &   J,EO,B    \\
131-046 &   5:35:13.05  &   -5:20:45.79 &   131-046 &   -   &   -   &   -   &   -   &   -   &   -   &   i   &         \\
131-247 &    5:35:13.11     &    -5:22:47.11    &   131-247 &   131-247 &   -   &   31  &   -   &   -   &   524 &   i   &   J     \\
132-1832    &    5:35:13.24     &    -5:18:32.95    &   -   &   132-1832    &   -   &   -   &   -   &   -   &   -   &   d   &   EO    \\
132-042 &    5:35:13.24     &    -5:20:41.94    &   132-042 &   132-042 &   -   &   36  &   -   &   -   &   -   &   i   &   J,EO      \\
133-353 &    5:35:13.30     &    -5:23:53.05    &   133-353 &   -   &   -   &   37  &   -   &   -   &   540 &   i   &   B     \\
135-220 &    5:35:13.51     &    -5:22:19.49    &   135-220 &   135-220 &   411 &   46  &   -   &   -   &   551 &   i   &         \\
138-207 &    5:35:13.78     &    -5:22:07.39    &   138-207 &   -   &   423 &   51  &   -   &   -   &   579 &   i   &         \\
139-320 &    5:35:13.92     &    -5:23:20.16    &   -   &   -   &   -   &   -   &   3430    &   -   &   593 &   i   &         \\
141-520 &    5:35:14.05     &    -5:25:20.50    &   141-520 &   141-520 &   -   &   -   &   2504    &   -   &   604 &   i   &   FO    \\
140-1952    &    5:35:14.05     &    -5:19:51.90    &   140-1952    &   141-1952    &   429 &   -   &   2364    &   J05351405-0519520   &   597 &   d   &   FO    \\
142-301 &    5:35:14.15     &    -5:23:00.91    &   142-301 &   141-301 &   -   &   -   &   2713    &   -   &   -   &   i   &         \\
143-425 &    5:35:14.27     &    -5:24:24.55    &   143-425 &   -   &   437 &   -   &   3100    &   J05351427-0524246   &   616 &   i   &         \\
144-522 &   5:35:14.34  &   -5:25:22.30 &   144-522 &   143-522 &   -   &   -   &   -   &   -   &   -   &   i   &         \\
146-201 &    5:35:14.61     &    -5:22:00.94    &   -   &   -   &   -   &   66  &   -   &   -   &   -   &   i   &         \\
147-323 &    5:35:14.72     &    -5:23:23.01    &   -   &   147-323 &   451 &   -   &   3527    &   -   &   658 &   i   &         \\
148-305 &    5:35:14.80     &    -5:23:04.76    &   148-305 &   -   &   -   &   73  &   -   &   -   &   664 &   i   &   B     \\
149-329 &    5:35:14.92     &    -5:23:29.05    &   149-329 &   -   &   455 &   -   &   -   &   -   &   671 &   i   &         \\
150-147 &    5:35:15.00     &    -5:21:47.34    &   -   &   -   &   -   &   83  &   -   &   -   &   -   &   i   &         \\
150-231 &    5:35:15.02     &    -5:22:31.11    &   150-231 &   -   &   -   &   86  &   -   &   -   &   678 &   i   &   B     \\
152-319 &    5:35:15.20     &    -5:23:18.81    &   152-319 &   -   &   -   &   -   &   3512    &   -   &   690 &   i   &         \\
152-738 &    5:35:15.21     &    -5:27:37.85    &   152-738 &   -   &   -   &   -   &   816 &   J05351521-0527378   &   693 &   i   &         \\
154-324 &    5:35:15.35     &    -5:23:24.11    &   154-324 &   -   &   -   &   -   &   -   &   -   &   -   &   i   &   B     \\
154-225 &    5:35:15.37     &    -5:22:25.35    &   154-225 &   -   &   472 &   96  &   -   &   -   &   699 &   i   &   B     \\
154-240 &    5:35:15.38     &    -5:22:39.85    &   -   &   154-240 &   -   &   -   &   -   &   -   &   -   &   i   &         \\
155-338 &    5:35:15.51     &    -5:23:37.45    &   155-338 &   155-338 &   -   &   -   &   3143    &   -   &   717 &   i   &         \\
157-323 &   5:35:15.72  &   -5:23:22.59 &   157-323 &   -   &   488 &   -   &   3253    &   -   &   733 &   i   &   B     \\
158-323 &   5:35:15.83  &   -5:23:22.59 &   158-323 &   -   &   -   &   -   &   3254    &   -   &   746 &   i   &   B     \\
158-327 &    5:35:15.79     &    -5:23:26.51    &   158-327 &   158-327 &   489 &   -   &   -   &   -   &   -   &   i   &   B     \\
158-326 &    5:35:15.81     &    -5:23:25.51    &   158-326 &   158-326 &   -   &   -   &   3254    &   -   &   747 &   i   &   B     \\
159-338 &    5:35:15.90     &    -5:23:38.00    &   159-338 &   -   &   -   &   -   &   -   &   -   &   757 &   i   &   B     \\
159-418 &    5:35:15.90     &    -5:24:17.85    &   159-418 &   159-418 &   -   &   118 &   3438    &   -   &   748 &   i   &         \\
159-221 &    5:35:15.93     &    -5:22:21.05    &   159-221 &   -   &   496 &   120 &   3456    &   -   &   756 &   d   &         \\
159-350 &    5:35:15.96     &    -5:23:50.30    &   159-350 &   -   &   499 &   -   &   3138    &   -   &   758 &   i   &   B     \\
160-353 &    5:35:16.01     &    -5:23:53.00    &   160-353 &   -   &   503 &   -   &   -   &   -   &   768 &   i   &   J     \\
161-324 &    5:35:16.05     &    -5:23:24.35    &   161-324 &   -   &   -   &   -   &   -   &   -   &   -   &   i   &         \\
161-328 &    5:35:16.07     &    -5:23:27.81    &   161-328 &   161-328 &   -   &   -   &   -   &   -   &   -   &   i   &         \\
161-314 &   5:35:16.10  &   -5:23:14.05 &   161-314 &   -   &   -   &   -   &   -   &   -   &   779 &   i   &         \\
162-133 &    5:35:16.19     &    -5:21:32.39    &   162-133 &   -   &   507 &   131 &   3352    &   -   &   783 &   i   &         \\
163-026 &   5:35:16.31  &   -5:20:25.24 &   -   &   163-026 &   510 &   137 &   2926    &   -   &   796 &   d   &   B     \\
163-210 &    5:35:16.27     &    -5:22:10.45    &   163-210 &   -   &   511 &   134 &   3177    &   -   &   784 &   i   &   B     \\
163-317 &    5:35:16.27     &    -5:23:16.51    &   163-317 &   -   &   512 &   -   &   -   &   -   &   787 &   i   &         \\
163-222 &    5:35:16.30     &    -5:22:21.50    &   163-222 &   163-222 &   -   &   140 &   -   &   -   &   799 &   i   &   B,FO      \\
163-249 &    5:35:16.33     &    -5:22:49.01    &   163-249 &   -   &   513 &   139 &   3167    &   -   &   800 &   i   &   B     \\
164-511 &   5:35:16.35  &   -5:25:09.60 &   164-511 &   -   &   516 &   141 &   2519    &   -   &   803 &   i   &         \\
165-235 &    5:35:16.48     &    -5:22:35.16    &   165-235 &   -   &   519 &   143 &   3518    &   -   &   807 &   i   &         \\
165-254 &    5:35:16.54     &    -5:22:53.70    &   -   &   165-254 &   -   &   -   &   -   &   -   &   -   &   d   &   RN    \\
166-316 &   5:35:16.61  &   -5:23:16.19 &   166-315 &   -   &   -   &   -   &   3528    &   -   &   820 &   i   &         \\
166-519 &    5:35:16.57     &    -5:25:17.74    &   166-519 &   -   &   -   &   147 &   3411    &   -   &   814 &   d   &         \\
166-406 &    5:35:16.57     &    -5:24:06.00    &   166-406 &   -   &   521 &   146 &   -   &   J05351675-0524041   &   813 &   i   &         \\
166-250 &    5:35:16.59     &    -5:22:50.36    &   166-250 &   -   &   -   &   -   &   -   &   -   &   -   &   i   &         \\
167-231 &    5:35:16.73     &    -5:22:31.30    &   167-231 &   167-231 &   -   &   151 &   3519    &   -   &   825 &   d   &   FO    \\
167-317 &   5:35:16.74  &   -5:23:16.51 &   167-317 &   -   &   524 &   -   &   -   &   -   &   826 &   i   &         \\
168-235 &   5:35:16.81  &   -5:22:34.71 &   168-235 &   -   &   -   &   -   &   -   &   -   &   -   &   i   &         \\
168-328 &    5:35:16.77     &    -5:23:28.06    &   168-328 &   -   &   -   &   -   &   -   &   -   &   827 &   i   &   B     \\
168-326 &    5:35:16.83     &    -5:23:25.91    &   168-326 &   -   &   -   &   -   &   -   &   -   &   -   &   i   &   B     \\
169-338 &    5:35:16.88     &    -5:23:38.10    &   169-338 &   -   &   -   &   -   &   -   &   -   &   -   &   i   &   B     \\
170-301 &    5:35:16.95     &    -5:23:00.91    &   170-301 &   -   &   533 &   161 &   3163    &   -   &   845 &   i   &         \\
170-249 &    5:35:16.96     &    -5:22:48.51    &   170-249 &   170-249 &   532 &   160 &   3260    &   -   &   844 &   i   &   B     \\
170-337 &    5:35:16.97     &    -5:23:37.15    &   170-337 &   170-337 &   534 &   -   &   -   &   -   &   847 &   i   &   B     \\
171-340 &    5:35:17.04     &    -5:23:39.75    &   171-340 &   171-340 &   537 &   -   &   3144    &   -   &   856 &   i   &   B     \\
171-434 &    5:35:17.11     &    -5:24:34.40    &   171-434 &   -   &   -   &   -   &   -   &   -   &   -   &   i   &   B     \\
172-028 &    5:35:17.22     &    -5:20:27.84    &   172-028 &   172-028 &   542 &   -   &   2921    &   -   &   865 &   d   &         \\
173-341 &    5:35:17.32     &    -5:23:41.40    &   173-341 &   -   &   -   &   178 &   -   &   -   &   886 &   i   &   B     \\
173-236 &    5:35:17.34     &    -5:22:35.81    &   173-236 &   174-236 &   548 &   174 &   -   &   -   &   876 &   i   &         \\
174-305 &    5:35:17.37     &    -5:23:04.86    &   174-305 &   -   &   -   &   175 &   3451    &   -   &   879 &   i   &         \\
174-414 &    5:35:17.38     &    -5:24:13.90    &   174-414 &   -   &   -   &   176 &   3419    &   -   &   887 &   i   &         \\
175-251 &    5:35:17.47     &    -5:22:51.26    &   175-251 &   -   &   -   &   181 &   -   &   -   &   884 &   i   &   B     \\
175-355 &   5:35:17.54  &   -5:23:55.05 &   175-355 &   175-355 &   -   &   -   &   -   &   -   &   -   &   i   &         \\
175-543 &    5:35:17.54     &    -5:25:42.89    &   175-543 &   176-543 &   557 &   182 &   -   &   -   &   901 &   i   &   J,EO,B    \\
176-325 &    5:35:17.55     &    -5:23:24.96    &   176-325 &   -   &   554 &   -   &   3529    &   -   &   900 &   i   &         \\
176-252 &    5:35:17.64     &    -5:22:51.66    &   176-252 &   -   &   -   &   188 &   -   &   -   &   906 &   i   &         \\
177-341W    &    5:35:17.66     &    -5:23:41.00    &   177-341 &   177-341 &   558 &   190 &   -   &   -   &   -   &   i   &   B     \\
177-454 &   5:35:17.69  &   -5:24:54.10 &   177-454 &   -   &   559 &   193 &   2544    &   -   &   914 &   i   &         \\
177-541 &    5:35:17.71     &    -5:25:40.76    &   -   &   177-541 &   -   &   -   &   -   &   -   &   -   &   i   &   EO,B      \\
177-444 &    5:35:17.73     &    -5:24:43.75    &   177-444 &   -   &   -   &   192 &   3407    &   -   &   -   &   i   &         \\
177-341E    &    5:35:17.73     &    -5:23:41.10    &   177-341 &   177-341 &   558 &   190 &   -   &   -   &   -   &   i   &   B     \\
178-441 &    5:35:17.81     &    -5:24:41.05    &   -   &   -   &   -   &   198 &   -   &   -   &   925 &   i   &         \\
178-258 &    5:35:17.84     &    -5:22:58.15    &   -   &   -   &   -   &   -   &   -   &   -   &   -   &   i   &         \\
179-056 &    5:35:17.92     &    -5:20:55.44    &   179-056 &   -   &   -   &   -   &   -   &   -   &   940 &   i   &   B     \\
179-354 &    5:35:17.96     &    -5:23:53.50    &   179-354 &   179-353 &   -   &   -   &   -   &   -   &   -   &   i   &         \\
180-331 &    5:35:18.03     &    -5:23:30.80    &   180-331 &   -   &   -   &   211 &   -   &   -   &   -   &   i   &   B     \\
181-247 &    5:35:18.08     &    -5:22:47.10    &   181-247 &   181-247 &   -   &   -   &   -   &   -   &   -   &   i   &   B     \\
181-825 &    5:35:18.10     &    -5:28:25.04    &   -   &   181-825 &   580 &   -   &   607 &   J05351810-0528249   &   948 &   i   &   RN    \\
182-316 &    5:35:18.19     &    -5:23:31.55    &   182-316 &   182-332 &   -   &   214 &   -   &   -   &   -   &   d   &   B     \\
182-413 &   5:35:18.22  &   -5:24:13.45 &   182-413 &   182-413 &   -   &   -   &   -   &   -   &   -   &   i   &         \\
183-439 &    5:35:18.28     &    -5:24:38.85    &   183-439 &   -   &   -   &   221 &   3298    &   -   &   -   &   i   &   B     \\
183-419 &   5:35:18.31  &   -5:24:18.85 &   183-419 &   183-419 &   -   &   -   &   -   &   -   &   -   &   i   &         \\
183-405 &    5:35:18.33     &    -5:24:04.85    &   183-405 &   183-405 &   588 &   233 &   -   &   -   &   966 &   d   &   FO,B      \\
184-427 &    5:35:18.35     &    -5:24:26.85    &   184-427 &   -   &   -   &   224 &   -   &   -   &   967 &   i   &   B     \\
184-520 &    5:35:18.44     &    -5:25:19.29    &   184-520 &   -   &   -   &   227 &   3410    &   -   &   -   &   i   &         \\
187-314 &    5:35:18.68     &    -5:23:14.01    &   187-314 &   -   &   596 &   233 &   -   &   -   &   986 &   i   &   B     \\
189-329 &   5:35:18.87  &   -5:23:28.85 &   189-329 &   -   &   604 &   240 &   3150    &   -   &   1000    &   i   &         \\
190-251 &    5:35:19.03     &    -5:22:50.65    &   -   &   -   &   -   &   -   &   -   &   -   &   -   &   i   &   B     \\
191-232 &   5:35:19.13  &   -5:22:31.20 &   -   &   191-232 &   -   &   -   &   -   &   -   &   -   &   d   &         \\
191-350 &    5:35:19.06     &    -5:23:49.50    &   191-350 &   191-350 &   607 &   244 &   3139    &   J05351906-0523495   &   1011    &   i   &   J     \\
193-1659    &    5:35:19.25     &    -5:16:58.69    &   -   &   -   &   -   &   -   &   -   &   -   &   -   &   rn  &         \\
197-427 &    5:35:19.65     &    -5:24:26.70    &   197-427 &   197-427 &   622 &   254 &   2594    &   -   &   1045    &   i   &   RN    \\
198-222 &    5:35:19.82     &    -5:22:21.55    &   198-222 &   -   &   624 &   255 &   -   &   -   &   1056    &   i   &         \\
198-448 &    5:35:19.83     &    -5:24:47.95    &   198-448 &   -   &   625 &   -   &   3305    &   -   &   1058    &   i   &         \\
199-1508    &    5:35:19.89     &    -5:15:08.25    &   -   &   -   &   -   &   -   &   2402    &   J05351983-0515089   &   1053    &   i   &   B     \\
200-106 &    5:35:20.04     &    -5:21:05.99    &   200-106 &   -   &   631 &   -   &   3194    &   J05352004-0521059   &   1071    &   i   &         \\
201-534 &   5:35:20.14  &   -5:25:33.84 &   201-534 &   -   &   -   &   260 &   -   &   -   &   -   &   i   &         \\
202-228 &    5:35:20.15     &    -5:22:28.30    &   202-228 &   -   &   -   &   261 &   -   &   -   &   1084    &   i   &   B     \\
203-504 &    5:35:20.26     &    -5:25:04.05    &   203-504 &   203-504 &   644 &   -   &   2530    &   -   &   1091    &   i   &   B     \\
203-506 &   5:35:20.32  &   -5:25:05.55 &   -   &   203-506 &   -   &   -   &   2530    &   -   &   -   &   d   &   RN    \\
205-330 &   5:35:20.45  &   -5:23:29.96 &   205-330 &   -   &   648 &   263 &   -   &   -   &   1101    &   i   &   B     \\
205-052 &    5:35:20.52     &    -5:20:52.05    &   205-052 &   -   &   650 &   -   &   2901    &   -   &   1104    &   i   &         \\
205-421 &    5:35:20.53     &    -5:24:21.00    &   205-421 &   205-421 &   652 &   265 &   3306    &   -   &   1107    &   i   &   FO    \\
206-446 &    5:35:20.62     &    -5:24:46.45    &   206-446 &   206-446 &   658 &   -   &   3097    &   -   &   1112    &   i   &         \\
208-122 &    5:35:20.83     &    -5:21:21.45    &   208-122 &   -   &   662 &   -   &   3351    &   -   &   1120    &   rn  &         \\
209-151 &    5:35:21.00     &    -5:21:52.30    &   209-151 &   -   &   665 &   -   &   -   &   -   &   1122    &   i   &   B     \\
210-225 &    5:35:21.03     &    -5:22:25.20    &   -   &   -   &   -   &   275 &   3491    &   -   &   -   &   i   &         \\
212-557 &    5:35:21.15     &    -5:25:57.04    &   212-557 &   -   &   674 &   -   &   3110    &   J05352115-0525569   &   1139    &   i   &         \\
212-400 &    5:35:21.19     &    -5:24:00.20    &   -   &   -   &   -   &   276 &   3443    &   -   &   -   &   i   &         \\
212-260 &    5:35:21.24     &    -5:22:59.51    &   212-260 &   -   &   -   &   280 &   3334    &   J05352124-0522594   &   1141    &   i   &         \\
213-533 &    5:35:21.28     &    -5:25:33.11    &   -   &   -   &   -   &   -   &   -   &   -   &   -   &   i   &   B     \\
213-346 &    5:35:21.30     &    -5:23:46.10    &   -   &   -   &   -   &   -   &   -   &   -   &   1149    &   i   &   B     \\
215-652 &    5:35:21.45     &    -5:26:52.40    &   -   &   -   &   -   &   -   &   -   &   -   &   -   &   i   &         \\
215-317 &    5:35:21.49     &    -5:23:16.71    &   215-317 &   -   &   685 &   284 &   3541    &   -   &   1155    &   i   &         \\
215-106 &    5:35:21.55     &    -5:21:05.60    &   215-106 &   -   &   684 &   -   &   2883    &   -   &   1154    &   i   &   J     \\
216-541 &    5:35:21.60     &    -5:25:40.70    &   -   &   -   &   -   &   -   &   2491    &   -   &   -   &   i   &         \\
216-715 &    5:35:21.62     &    -5:27:14.65    &   216-715 &   -   &   689 &   -   &   825 &   J05352162-0527145   &   1163    &   i   &         \\
218-339 &    5:35:21.77     &    -5:23:39.30    &   218-339 &   -   &   694 &   289 &   3146    &   J05352177-0523392   &   1167    &   i   &         \\
218-354 &    5:35:21.79     &    -5:23:53.90    &   218-354 &   218-354 &   698 &   290 &   3332    &   J05352181-0523539   &   1174    &   d   &   EO,B      \\
218-529 &    5:35:21.82     &    -5:25:28.46    &   -   &   218-529 &   -   &   -   &   -   &   -   &   -   &   i   &   B     \\
218-306 &    5:35:21.84     &    -5:23:06.46    &   -   &   -   &   -   &   294 &   3542    &   -   &   1173    &   i   &   B     \\
221-433 &    5:35:22.08     &    -5:24:32.95    &   221-433 &   -   &   -   &   -   &   3099    &   -   &   1184    &   i   &         \\
223-414 &    5:35:22.31     &    -5:24:14.25    &   223-414 &   -   &   710 &   -   &   3124    &   J05352232-0524141   &   1205    &   i   &   B     \\
224-728 &    5:35:22.37     &    -5:27:28.40    &   224-728 &   -   &   716 &   -   &   824 &   J05352237-0527283   &   1206    &   i   &         \\
228-548 &    5:35:22.83     &    -5:25:47.69    &   228-548 &   -   &   724 &   -   &   2487    &   -   &   -   &   i   &         \\
230-536 &    5:35:23.02     &    -5:25:36.29    &   -   &   -   &   -   &   -   &   3507    &   -   &   -   &   d   &         \\
231-460 &   5:35:23.05  &   -5:24:59.86  &   231-460  &   -   &   -   &   -   &   -   &   -   &   1237    &   i   &         \\
231-502 &   5:35:23.16  &   -5:25:02.19  &   231-502  &   -   &   -   &   -   &   -   &   -   &   -   &   i   &   B     \\
231-838 &    5:35:23.10    &    -5:28:37.34    &  231-838  &  -  &  -  &   -   &   -   &   -   &   1238    &   i  &  J   \\
232-453 &   5:35:23.22  &   -5:24:52.79 &   232-453 &   -   &   -   &   -   &   -   &   -   &   -   &   i   &         \\
234-853 &    5:35:23.40     &    -5:28:53.19    &   -   &   -   &   -   &   -   &   -   &   -   &   -   &   i   &         \\
236-527 &    5:35:23.59     &    -5:25:26.54    &   236-527 &   -   &   737 &   -   &   3207    &   -   &   1262    &   i   &         \\
237-627 &    5:35:23.66     &    -5:26:27.15    &   237-627 &   -   &   743 &   -   &   3073    &   -   &   1263    &   i   &         \\
238-334 &    5:35:23.80     &    -5:23:34.30    &   238-334 &   -   &   744 &   -   &   3152    &   -   &   1268    &   i   &   B     \\
239-334 &    5:35:23.86     &    -5:23:34.05    &   -   &   239-334 &   -   &   -   &   -   &   -   &   -   &   d   &   EO,B      \\
239-510 &    5:35:23.98     &    -5:25:09.94    &   239-510 &   -   &   -   &   -   &   3319    &   -   &   1275    &   i   &         \\
240-314 &    5:35:24.02     &    -5:23:13.85    &   240-314 &   -   &   -   &   -   &   3502    &   J05352402-0523138   &   1276    &   i   &   B     \\
242-519 &    5:35:24.22     &    -5:25:18.79    &   242-519 &   242-519 &   750 &   -   &   2516    &   J05352425-0525186   &   1281    &   i   &         \\
244-440 &    5:35:24.38     &    -5:24:39.74    &   244-440 &   244-440 &   756 &   -   &   3098    &   J05352443-0524398   &   1290    &   i   &         \\
245-632 &    5:35:24.45     &    -5:26:31.55    &   245-632 &   -   &   758 &   -   &   3060    &   J05352445-0526314   &   1291    &   i   &   B     \\
245-1910    &    5:35:24.48     &    -5:19:09.84    &   -   &   -   &   -   &   -   &   2350    &   -   &   -   &   i   &   RN,B      \\
245-502 &    5:35:24.51     &    -5:25:01.59    &   245-502 &   -   &   759 &   -   &   2541    &   -   &   1293    &   i   &         \\
247-436 &    5:35:24.69     &    -5:24:35.74    &   247-436 &   -   &   762 &   -   &   3087    &   -   &   1302    &   i   &   J     \\
250-439 &    5:35:25.02     &    -5:24:38.49    &   250-439 &   -   &   -   &   -   &   -   &   -   &   1313    &   i   &         \\
252-457 &    5:35:25.21     &    -5:24:57.34    &   252-457 &   252-457 &   773 &   -   &   2553    &   -   &   1317    &   i   &         \\
253-1536    &    5:35:25.30     &    -5:15:35.54    &   -   &   253-1536    &   -   &   -   &   -   &   -   &   -   &   d   &   J,RN,B    \\
254-412 &    5:35:25.37     &    -5:24:11.50    &   254-412 &   -   &   775 &   -   &   3128    &   -   &   1323    &   i   &         \\
255-512 &    5:35:25.52     &    -5:25:11.84    &   -   &   -   &   -   &   -   &   -   &   -   &   -   &   i   &         \\
262-521 &    5:35:26.18     &    -5:25:20.49    &   262-521 &   -   &   -   &   -   &   3321    &   -   &   1345    &   i   &         \\
264-532 &    5:35:26.42     &    -5:25:31.69    &   264-532 &   -   &   -   &   -   &   3505    &   -   &   -   &   i   &         \\
266-558 &    5:35:26.62 &    -5:25:57.84    &   -   &   266-558 &   -   &   -   &   -   &   -   &   -   &   i   &         \\
280-931 &    5:35:27.96     &    -5:29:31.15    &   -   &   -   &   824 &   -   &   2276    &   J05352797-0529311   &   1403    &   i   &         \\
280-1720    &    5:35:28.05     &    -5:17:20.33    &   -   &   -   &   821 &   -   &   1314    &   J05352804-0517202   &   1404    &   d   &   FO    \\
281-306 &    5:35:28.13     &    -5:23:06.45    &   -   &   -   &   825 &   -   &   3164    &   J05352813-0523064   &   1407    &   d   &   FO    \\
282-458 &    5:35:28.20     &    -5:24:58.19    &   282-458 &   282-458 &   826 &   -   &   3080    &   -   &   1409    &   i   &   J     \\
282-614 &    5:35:28.20     &    -5:26:14.20    &   -   &   -   &   -   &   -   &   -   &   -   &   -   &   i   &         \\
284-439 &    5:35:28.40     &    -5:24:38.69    &   -   &   -   &   -   &   -   &   -   &   J05352840-0524386   &   1414    &   i   &         \\
294-757 &    5:35:29.43     &    -5:37:56.60    &   -   &   -   &   -   &   -   &   -   &   J05352943-0537563   &   -   &   rn  &         \\
294-606 &    5:35:29.48     &    -5:26:06.63    &   -   &   294-606 &   -   &   -   &   -   &   -   &   -   &   d   &   RN    \\
297-025 &    5:35:29.67     &    -5:30:24.75    &   -   &   -   &   -   &   -   &   622 &   J05352967-0530247   &   1431    &   i   &         \\
304-539 &    5:35:30.41     &    -5:25:38.63    &   304-539 &   -   &   850 &   -   &   2333    &   J05353042-0525385   &   1444    &   i   &         \\
307-1807    &   5:35:30.70  &   -5:18:07.24 &   307-1807    &   -   &   854 &   -   &   -   &   J05353070-0518071   &   1449    &   i   &         \\
314-816 &    5:35:31.40     &    -5:28:16.48    &   -   &   -   &   872 &   -   &   755 &   J05353141-0528163   &   1474    &   i   &         \\
321-602 &    5:35:32.10     &    -5:26:01.94    &   321-602 &   -   &   -   &   -   &   -   &   -   &   -   &   d   &   EO    \\
332-405 &    5:35:33.19     &    -5:24:04.74    &   -   &   -   &   -   &   -   &   -   &   J05353316-0524050   &   -   &   d   &         \\
332-1605    &   5:35:33.20  &   -5:16:05.38 &   332-1605    &   -   &   -   &   -   &   2411    &   J05353319-0516053   &   -   &   i   &         \\
346-1553    &    5:35:34.62     &    -5:15:52.92    &   -   &   -   &   903 &   -   &   1701    &   -   &   -   &   d   &   FO    \\
347-1535    &   5:35:34.67  &   -5:15:34.88 &   347-1535    &   -   &   -   &   -   &   -   &   -   &   -   &   d   &   J     \\
351-3349    &    5:35:35.13     &    -5:33:49.18    &   -   &   -   &   913 &   -   &   306 &   -   &   -   &   i   &   J     \\
353-130 &    5:35:35.32     &    -5:21:29.59    &   -   &   -   &   -   &   -   &   -   &   -   &   -   &   j   &   J,B   \\
473-245 &    5:35:47.34     &    -5:22:44.82    &   -   &   -   &   -   &   -   &   -   &   -   &   -   &   d   &   EO    \\

\enddata
\tablenotetext{a}{Units of right ascension are hours, minutes,
seconds and units of declination are degrees, arcminutes and
arcseconds (J2000.0).} \tablenotetext{b}{The abbreviations of the
catalogues are: OW-O'Dell \& Wen 1994, O'Dell \& Wong 1996 and
O'Dell 2001; BOM-Bally et al. 2000 and Smith et al. 2005;
P-Prosser et al. 1994; JW-Jones \& Walker 1988; AD-Ali \& DePoy
1995; 2MASS-2Micron All-Sky Survey (Cutri et al. 2003);
COUP-Chandra Orion Ultradeep Project (Getman et al. 2005).}
\tablenotetext{c}{In this column, i: ionized disk seen in
emission; d: dark disk seen only in silhouette; rn: reflection
nebulae with no external ionized gas emission; j: jet emission
with no evidence of neither ionized disk nor silhouette disk.}
\tablenotetext{d}{In this column, J: jet; RN: reflection nebula;
EO: disk seen nearly edge-on; FO: disk seen nearly face-on; B:
binary system.}

\end{deluxetable}

\clearpage
\begin{deluxetable}{lcccccccccc}
\tabletypesize{\scriptsize} \rotate
\renewcommand{\arraystretch}{0.6}
\tablecaption{Previously observed proto-planetary disks not
detected from the HST/ACS images.\label{other_disks_tab}}
\tablewidth{0pt} \tablehead{ \colhead{Object} &
\colhead{RA\tablenotemark{a}} & \colhead{DEC\tablenotemark{a}} &
\colhead{OW\tablenotemark{b}} & \colhead{BOM\tablenotemark{b}} &
\colhead{JW\tablenotemark{b}}& \colhead{P\tablenotemark{b}}&
\colhead{AD\tablenotemark{b}}& \colhead{2MASS\tablenotemark{b}}&
\colhead{COUP\tablenotemark{b}}& \colhead{Note\tablenotemark{c}}}
\startdata

158-314 &   5:35:15.82  &   -5:23:14.56 &   158-314 &   -   &   -   &   -   &   -   &   -   &   745 &  close to bright star    \\
163-322 &   5:35:16.29  &   -5:23:21.76 &   163-322 &   -   &   -   &   -   &   -   &   -   &   -   &  close to bright star    \\
163-323 &   5:35:16.32  &   -5:23:22.67 &   163-323 &   -   &   -   &   -   &   -   &   -   &   -   &  close to bright star    \\
216-0939    & 5:35:21.57   &  -5:09:38.9   &   -   &   216-0939    &   676 &   -   &   1914    &   -   &   -   & out of ACS FOV      \\

\enddata

\tablenotetext{a}{Units of right ascension are hours, minutes,
seconds and units of declination are degrees, arcminutes and
arcseconds (J2000.0). Since the objects in this table could not be
found in our images we kept their previously reported
coordinates.} \tablenotetext{b}{The abbreviations of the
catalogues are: OW-O'Dell \& Wen 1994, O'Dell \& Wong 1996 and
O'Dell 2001; BOM-Bally et al. 2000 and Smith et al. 2005;
P-Prosser et al. 1994; JW-Jones \& Walker 1988; AD-Ali \& DePoy
1995; 2MASS-2Micron All-Sky Survey (Cutri et al. 2003);
COUP-Chandra Orion Ultradeep Project (Getman et al. 2005).}
\tablenotetext{c}{This columns explains the reason why we could
not detect the objects listed in this table: close to bright
star-under the saturation bleeding trail of a close bright star;
out of ACS FOV-out of HST/ACS Treasury Program field of view.}

\end{deluxetable}

\clearpage
\begin{deluxetable}{lcccccccccc}
\tabletypesize{\scriptsize} \rotate
\renewcommand{\arraystretch}{0.6}
\tablecaption{Previously observed extended objects not identified
as circumstellar disks from the HST/ACS
images.\label{OW_BOM_objects_tab}} \tablewidth{0pt} \tablehead{
\colhead{Object} & \colhead{RA\tablenotemark{a}} &
\colhead{DEC\tablenotemark{a}} & \colhead{OW\tablenotemark{b}} &
\colhead{BOM\tablenotemark{b}} & \colhead{JW\tablenotemark{b}} &
\colhead{P\tablenotemark{b}} & \colhead{AD\tablenotemark{b}} &
\colhead{2MASS\tablenotemark{b}} & \colhead{COUP\tablenotemark{b}}
& \colhead{ACS Type\tablenotemark{c}}} \startdata

064-705 &   5:35:06.43  &   -5:27:04.74 &   064-705 &   -   &   290 &   -   &   -   &   J05350642-0527048   &   266 &   B   \\
097-125 &   5:35:09.69  &   -5:21:24.89 &   097-125 &   -   &   332 &   -   &   3187    &   -   &   336 &   *   \\
109-449 &   5:35:10.93  &   -5:24:48.65 &   109-449 &   -   &   356 &   -   &   3103    &   J05351094-0524486   &   404 &   *   \\
113-153 &   5:35:11.31  &   -5:21:53.14 &   113-153 &   -   &   -   &   -   &   2801    &   -   &   -   &   HH  \\
114-155 &   5:35:11.34  &   -5:21:53.94 &   114-155 &   -   &   -   &   -   &   2801    &   -   &   -   &   HH  \\
115-155 &   5:35:11.50  &   -5:21:54.29 &   115-155 &   -   &   -   &   -   &   -   &   -   &   -   &   HH  \\
116-156 &   5:35:11.54  &   -5:21:55.64 &   116-156 &   -   &   -   &   -   &   -   &   -   &   -   &   HH  \\
117-025 &   5:35:11.70  &   -5:20:25.19 &   117-025 &   -   &   -   &   -   &   2920    &   -   &   -   &   HH  \\
127-711 &   5:35:12.70  &   -5:27:10.75 &   127-711 &   -   &   392 &   -   &   3008    &   J05351270-0527106   &   498 &   B   \\
128-044 &   5:35:12.81  &   -5:20:43.68 &   128-044 &   -   &   391 &   23  &   3239    &   J05351281-0520436   &   501 &   *   \\
132-221 &   5:35:13.17  &   -5:22:20.89 &   132-221 &   -   &   399 &   33  &   3522    &   -   &   523 &   B   \\
132-222 &   5:35:13.17  &   -5:22:21.09 &   132-222 &   -   &   -   &   -   &   -   &   -   &   -   &   B   \\
135-227 &   5:35:13.35  &   -5:22:26.11 &   135-227 &   -   &   404 &   41  &   3523    &       &   538 &   *   \\
137-222 &   5:35:13.72  &   -5:22:22.19 &   137-222 &   -   &   420 &   50  &   -   &   -   &   573 &   *   \\
144-334 &   5:35:14.39  &   -5:23:33.50 &   144-334 &   -   &   441 &   -   &   -   &   -   &   631 &   *   \\
153-321 &   5:35:15.35  &   -5:23:21.25 &   153-321 &   -   &   -   &   -   &   -   &   -   &   -   &   *    \\
153-1902 &   5:35:15.35  &   -5:19:02.15 &   153-1902    &   -   &   469 &   -   &   1374    &   J05351534-0519021   &   695 &   *    \\
154-042 &   5:35:15.48  &   -5:20:42.14 &   154-042 &   -   &   -   &   -   &   2903    &   -   &   -   &   HH  \\
155-040 &   5:35:15.47  &   -5:20:40.25 &   155-040 &   -   &   -   &   -   &   2907    &   -   &   703 &   HH  \\
156-403 &   5:35:15.61  &   -5:24:03.15 &   156-403 &   -   &   480 &   105 &   3125    &   -   &   726 &   *   \\
158-425 &   5:35:15.77  &   -5:24:24.75 &   158-425 &   -   &   -   &   112 &   3403    &   -   &   736 &   B   \\
168-309 &   5:35:16.81  &   -5:23:09.93 &   168-309 &   -   &   -   &   -   &   -   &   -   &   -   &       \\
171-315 &   5:35:17.88  &   -5:23:15.48 &   171-315 &   -   &   -   &   -   &   -   &   -   &   -   &   B   \\
171-334 &   5:35:17.06  &   -5:23:33.95 &   171-334 &   -   &   538 &   -   &   -   &   J05351705-0523341   &   855 &   *   \\
172-327 &   5:35:17.20  &   -5:23:26.66 &   172-327 &   -   &   -   &   -   &   -   &   -   &   -   &   ISM \\
174-400 &   5:35:17.38  &   -5:24:00.25 &   174-400 &   -   &   -   &   -   &   -   &   -   &   -   &   *   \\
179-534 &   5:35:17.94  &   -5:25:33.79 &   174-534 &   -   &   570 &   203 &   3303    &   -   &   937 &   B   \\
179-536 &   5:35:17.88  &   -5:25:36.03 &   174-536 &   -   &   -   &   -   &   2495    &   -   &   -   &       \\
201-534 &   5:35:20.14  &   -5:25:33.84 &   201-534 &   -   &   -   &   260 &   -   &   -   &   -   &   *   \\
222-637 &   5:35:22.18  &   -5:26:37.40 &   222-637 &   -   &   709 &   -   &   3074    &   J05352219-0526373   &   1202    &   B   \\

\enddata

\tablenotetext{a}{Units of right ascension are hours, minutes,
seconds and units of declination are degrees, arcminutes and
arcseconds (J2000.0). When no objects could be found in our images
around a $2''$-radius circle from their previously reported
coordinates, we kept these latter ones.} \tablenotetext{b}{The
abbreviations of the catalogues are: OW-O'Dell \& Wen 1994, O'Dell
\& Wong 1996 and O'Dell 2001; BOM-Bally et al. 2000 and Smith et
al. 2005; P-Prosser et al. 1994; JW-Jones \& Walker 1988; AD-Ali
\& DePoy 1995; 2MASS-2Micron All-Sky Survey (Cutri et al. 2003);
COUP-Chandra Orion Ultradeep Project (Getman et al. 2005).}
\tablenotetext{c}{Object type as it appears from HST/ACS images:
B-close binary system; HH-Herbig-Haro object; *-star;
ISM-interstellar material.}

\end{deluxetable}

\clearpage
\begin{deluxetable}{lccccccccc}
\tabletypesize{\scriptsize} \rotate
\renewcommand{\arraystretch}{0.6}
\tablecaption{Extended objects detected from the HST/ACS Treasury
Program described in \S 7.\label{other_objects_tab}}
\tablewidth{0pt} \tablehead{ \colhead{Object} &
\colhead{RA\tablenotemark{a}} & \colhead{DEC\tablenotemark{a}} &
\colhead{OW\tablenotemark{b}} & \colhead{BOM\tablenotemark{b}} &
\colhead{JW\tablenotemark{b}}& \colhead{P\tablenotemark{b}}&
\colhead{AD\tablenotemark{b}}& \colhead{2MASS\tablenotemark{b}}&
\colhead{COUP\tablenotemark{b}}} \startdata

4222-1559   &    5:34:22.15     &    -5:15:58.94    &   -   &   -   &   -   &   -   &   -   &   -   &   -          \\
4305-1410   &    5:34:30.48     &    -5:14:09.79    &   -   &   -   &   -   &   -   &   -   &   -   &   -           \\
4407-1258   &    5:34:40.67     &    -5:12:57.71    &   -   &   -   &   -   &   -   &   -   &   -   &   -           \\
4422-3641   &    5:34:42.19     &    -5:36:41.02    &   -   &   -   &   -   &   -   &   -   &   -   &   -           \\
4425-228    &    5:34:42.48     &    -5:22:27.75    &   -   &   -   &   -   &   -   &   999 &   -   &   -           \\
222-3155    &    5:35:22.23     &    -5:35:15.45    &   -   &   -   &   -   &   -   &   -   &   -   &   -          \\
228-3918    &    5:35:22.79     &    -5:39:18.45    &   -   &   -   &   -   &   -   &   -   &   -   &   -          \\
295-753 &    5:35:29.45     &    -5:37:53.25    &   -   &   -   &   -   &   -   &   -   &   -   &   -             \\
297-334 &    5:35:29.69     &    -5:33:33.74    &   -   &   -   &   -   &   -   &   -   &   -   &   -           \\
299-309 &    5:35:29.92     &    -5:33:08.69    &   -   &   -   &   -   &   -   &   -   &   -   &   -           \\
311-3744    &    5:35:31.11     &    -5:37:44.39    &   -   &   -   &   -   &   -   &   -   &   -   &   -           \\
320-3938    &    5:35:32.03     &    -5:39:38.45    &   -   &   -   &   -   &   -   &   -   &   -   &   -         \\
355-3607    &    5:35:35.47     &    -5:36:06.77    &   -   &   -   &   -   &   -   &   -   &   -   &   -         \\
359-3141    &    5:35:35.94     &    -5:31:41.10    &   -   &   -   &   -   &   -   &   -   &   -   &   -         \\
364-3425    &    5:35:36.43     &    -5:34:25.10    &   -   &   -   &   -   &   -   &   -   &   -   &   -           \\
440-3246    &    5:35:44.01     &    -5:32:46.34    &   -   &   -   &   -   &   -   &   -   &   -   &   -        \\

\enddata
\tablenotetext{a}{Units of right ascension are hours, minutes,
seconds and units of declination are degrees, arcminutes and
arcseconds (J2000.0).} \tablenotetext{b}{The abbreviations of the
catalogues are: OW-O'Dell \& Wen 1994, O'Dell \& Wong 1996 and
O'Dell 2001; BOM-Bally et al. 2000 and Smith et al. 2005;
P-Prosser et al. 1994; JW-Jones \& Walker 1988; AD-Ali \& DePoy
1995; 2MASS-2Micron All-Sky Survey (Cutri et al. 2003);
COUP-Chandra Orion Ultradeep Project (Getman et al. 2005).}

\end{deluxetable}

\clearpage

\end{document}